%% file: inflation.tex
\def\isweb{1}
\documentclass{pdg}

\reviewtitle{Inflation}
\reviewauthor{J. Ellis (King's Coll. London; CERN), V. Vennin (LPENS; CNRS) and D. Wands (Portsmouth U.)}
\reviewlabel{inflation}
\ischaptertrue


\usepackage{makeidx}
\makeindex

\ifdefined\isbooklet
\externalcitedocument{inflation-full}
\fi

\IfFileExists{crossref.aux}{\externaldocument{crossref}}{}

\begin{document}

\ifdefined\isbook
\setcounter{page}{457}
\fi
\ifdefined\isbooklet
\setcounter{page}{234}
\fi

\ifdefined\isbook
\rppthumb
\fi

\ifdefined\isbooklet
\fontsize{8pt}{9pt}\selectfont
\input{inflation-booklet}
\else
\begin{bibunit}
\input{inflation-main}

\end{bibunit}
\fi

\ifdefined\isdraft
\clearpage
\renewcommand{\twocolumn}[1][]{
     \twocolumngrid
     #1
}
\printindex
\fi

\end{document}

%% file: inflation-main.tex
%
%

%
\pdgtitle


\revised{August 2025}

\section{Motivation and Introduction}

%


\def\R{{\mathrm{R}}}

\index{Inflation!motivation}%
\index{Universe!thermal history of}%
The standard Big-Bang model of cosmology provides a successful framework in which to understand 
the thermal history of our Universe and the growth of cosmic structure, but it is essentially 
incomplete. As described in Sec{bigbang:sec:problems},
\index{Cosmology!Big-Bang}%
\index{Big-Bang cosmology}%
Big-Bang cosmology requires very specific initial conditions. It postulates a uniform cosmological 
background, described by a spatially-flat, homogeneous and isotropic 
\index{Robertson-Walker metric}%
\index{Cosmology!Robertson-Walker metric}%
Robertson-Walker (RW) metric (\Eq{bigbang:eq:RWmetric} in ``Big-Bang Cosmology'' review), with 
scale factor $R(t)$. Within this setting, it also requires an initial almost scale-invariant 
distribution of primordial density perturbations as seen, for example, in the cosmic microwave 
background (CMB) radiation (described in Chap. \ref{microwave}, ``Cosmic Microwave Background'' review), 
on scales far larger than the causal horizon at the time the CMB photons last scattered.
\index{Inflation!CMB photons}%

The Hubble expansion rate, $H\equiv\dot{R}/R$, in an RW cosmology is given by the 
\index{Friedmann constraint equation}%
\index{Inflation!Friedman equation}%
Friedmann constraint equation (\Eq{bigbang:eq:ee00} in ``Big-Bang Cosmology'' review)
\begin{equation}\label{inflation:eq:FriedmannInf}
H^2 = {\frac{8\pi \rho}{3 M_{\rm P}^2}} + {\frac{{\rm \Lambda}}{ 3}} - {\frac{k}{ R^2}} \,,
\end{equation}
where $k/R^2$ is the intrinsic spatial curvature.
We use natural units such that the speed of light $c=1$ and hence we have the 
Planck mass\footnote{We note that another commonly used convention in the literature ({\textit{e.g.}}, Refs.~\cite{Lyth:1998xn,inflation:liddle/lyth:CIALSS,Baumann:2009ds,Martin:2013tda} is to work in terms of the reduced Planck mass, $(8\pi G_{\rm N})^{-1/2}\simeq 2\times10^{18}$~GeV.} $M_{\rm P}=G_{\rm N}^{-1/2}\simeq 10^{19}$~GeV
(see ``Astrophysical Constants and Parameters'').
\index{Cosmological!constant ${\rm \Lambda}$}%
A cosmological constant, ${\rm \Lambda}$, of the magnitude required to accelerate the Universe 
today (see Chap. \ref{darkenergy}, ``Dark Energy'' review) would have been completely negligible 
in the early Universe where the energy density $\rho\gg M_{\rm P}^2{\rm \Lambda}\sim 10^{-12}({\textrm{eV}})^4$. 
\index{Cosmology!standard early Universe}%
The standard early Universe cosmology, described in \Sec{bigbang:sec:SMsolusions} 
in ``Big-Bang Cosmology'' review, is thus dominated by non-relativistic matter 
($p_{\rm m}\ll \rho_{\rm m}$) 
or radiation ($p_{\rm r}=\rho_{\rm r}/3$ for an isotropic distribution). This leads to a decelerating expansion with $\ddot{R}<0$.

\index{Inflation!of early Universe}%
\index{Early Universe}%
The hypothesis of inflation~\cite{Starobinsky:1980te,Guth:1980zm} postulates a 
period of accelerated expansion, $\ddot{R}>0$, in the very early Universe, 
preceding the standard radiation-dominated era, 
which offers a physical model for the origin of these initial conditions, as 
reviewed in Refs.~\cite{Olive:1989nu,Lyth:1998xn,inflation:liddle/lyth:CIALSS,Baumann:2009ds,Martin:2013tda,Martin:2013nzq,Martin:2015dha}.
Such a period of accelerated expansion (i) drives a curved Robertson-Walker 
spacetime (with spherical or hyperbolic spatial geometry) towards spatial 
flatness, and (ii) it also expands the causal horizon beyond the present 
Hubble length, so as to encompass all the scales relevant to describe the 
large-scale structure observed in our Universe today, via the following two mechanisms.

\begin{enumerate}
\item A spatially-flat Universe with vanishing spatial curvature, $k=0$, has the dimensionless density parameter ${\rm \Omega}_{\rm tot}=1$,
where we define (\Eq{bigbang:eq:omo} in ``Big-Bang Cosmology'' review; 
see Chap. \ref{hubble}, ``Cosmological Parameters'' review for more complete definitions)
\begin{equation}\label{inflation:eq:eq1}
{\rm \Omega}_{\rm tot} 
\equiv {\frac{8\pi\rho_{\rm tot}}{ 3M_{\rm P}^2 H^2}} \,,
\end{equation}
with $\rho_{\rm tot} \equiv \rho + {\rm \Lambda} M_{\rm P}^2/8 \pi$.
If we re-write the Friedmann constraint (\Eq{inflation:eq:FriedmannInf}) in terms of ${\rm \Omega}_{\rm tot}$ we have
\begin{equation}\label{inflation:eq:FriedmannOmega}
1- {\rm \Omega}_{\rm tot} =  - {\frac{k}{ \dot{R}^2}} \,.
\end{equation}
Observations require $|1-{\rm \Omega}_{{\rm tot},0}|<0.005$ today\cite{Ade:2015xua}, where the subscript "$0$" denotes the present-day value.
Taking the time derivative of \Eq{inflation:eq:FriedmannOmega} we obtain
\begin{equation}\label{inflation:eq:eq2}
{\frac{d}{dt}} \left( 1- {\rm \Omega}_{\rm tot} \right) = - 2 {\frac{\ddot{R}}{\dot{R}}} \left( 1 - {\rm \Omega}_{\rm tot} \right) \,.
\end{equation}
Thus in a decelerating expansion, $\dot{R}>0$ and $\ddot{R}<0$,
any small initial deviation from spatial flatness grows, $(d/dt)|1-{\rm \Omega}_{\rm tot}|>0$. 
A small value such as $|1-{\rm \Omega}_{{\rm tot},0}|<0.005$ today requires an even smaller 
value at earlier times, {\textit{e.g.}}, $|1-{\rm \Omega}_{{\rm tot}}|<10^{-5}$ at the last scattering 
of the CMB, which appears unlikely, unless for some reason space is exactly flat.
However, an extended period of accelerated expansion in the very early Universe, with $\dot{R}>0$ and $\ddot{R}>0$
and hence $(d/dt)|1-{\rm \Omega}_{\rm tot}|<0$, can drive ${\rm \Omega}_{\rm tot}$ sufficiently 
close to unity, so that $|1-{\rm \Omega}_{{\rm tot},0}|$ remains unobservably small today, 
even after the radiation- and matter-dominated eras, for a wide range of initial values of ${\rm \Omega}_{\rm tot}$.

\item
The comoving distance (the present-day proper distance) traversed by light between 
cosmic time $t_1$ and $t_2$ in an expanding
Universe can be written, (see \Eq{bigbang:eq:deltar} in ``Big-Bang Cosmology'' review), as
\begin{equation}\label{inflation:eq:D0}
D_0(t_1,t_2) 
 = R_0 \int_{t_1}^{t_2} {\frac{dt}{R(t)}} 
= R_0 \int_{\ln R_1}^{\ln R_2} {\frac{d(\ln R)}{\dot{R}}} \,.
\end{equation}
In standard decelerated
(radiation- or matter-dominated)
cosmology the integrand, $1/\dot{R}$,
decreases towards the past, and there is a finite comoving distance traversed by 
light (a particle horizon) since the Big Bang ($R_1\to0$).
For example, the comoving size of the particle horizon at the CMB last-scattering 
surface ($R_2=R_{\rm lss}$) corresponds to $D_0\simeq100$\,Mpc, or approximately 
$1^\circ$ on the CMB sky today (see \Sec{bigbang:sec:problems} 
in ``Big-Bang Cosmology'' review).
However, during a period of inflation, $1/\dot{R}$ increases towards the past, 
and hence the integral (\Eq{inflation:eq:D0}) diverges as $R_1\to0$, allowing an 
arbitrarily large causal horizon, dependent only upon the duration of the 
accelerated expansion.
Assuming that the Universe inflates with a finite Hubble rate $H_*$ at $t_1=t_*$, 
ending with $H_{\rm end}<H_*$ at $t_2=t_{\rm end}$, we have
\begin{equation}\label{inflation:eq:eq3}
D_0(t_*,t_{\rm end})
> 
\left( {\frac{R_0}{ R_{\rm end}}} \right)
H_*^{-1} 
\left( e^{N_*} -1 \right) \,,
\end{equation}
where $N_*\equiv \ln (R_{\rm end} /R_*)$ describes the duration of inflation, 
measured in terms of the logarithmic expansion (or ``e-folds'') from $t_1=t_*$ 
up to the end of inflation at $t_2=t_{\rm end}$, and $R_0/R_{\rm end}$ is the 
subsequent expansion from the end of inflation to the present day. If inflation 
occurs above the TeV scale, the comoving Hubble scale at the end of inflation, 
$(R_0/R_{\rm end}) H_{\rm end}^{-1}$, is less than one astronomical 
unit ($\sim 10^{11}$~m), and a causally-connected patch can encompass our 
entire observable Universe today, which has a size $D_0>30$~Gpc, if there 
were more than 40 e-folds of inflation ($N_*>40$).
If inflation occurs at the GUT scale ($10^{15}$~GeV) then we require more than 60 e-folds.
\end{enumerate}

Producing an accelerated expansion in general relativity requires an 
energy-momentum tensor with negative pressure, $p<-\rho/3$ (see \Eq{bigbang:eq:eeii} 
in ``Big-Bang Cosmology'' review and Chap. \ref{darkenergy}, ``Dark Energy'' review), 
quite different from the hot dense plasma of relativistic particles in the hot Big Bang.
However a positive vacuum energy $V > 0$ does exert a negative pressure, $p_V=-\rho_V$. 
The work done by the cosmological expansion must be negative in this case so that the 
local vacuum energy density remains constant in an expanding
Universe, $\dot\rho_V=-3H(\rho_V+p_V)=0$. Therefore, a false vacuum state can drive an 
exponential expansion, corresponding to a de Sitter spacetime with a constant Hubble 
rate $H^2=8\pi\rho_V/3M_{\rm P}^2$ on spatially-flat hypersurfaces.

A constant vacuum energy $V$, equivalent to a cosmological constant ${\rm \Lambda}$ in the Friedmann equation
\Eq{inflation:eq:FriedmannInf}, cannot provide a complete description of inflation in 
the early Universe, since inflation must necessarily have come to an end in order for 
the standard Big-Bang cosmology to follow. A phase transition to the present true 
vacuum is required to release the false vacuum energy into the energetic plasma of 
the hot Big Bang and produce the large total entropy of our observed Universe today. 
Thus, we must necessarily study dynamical models of inflation, where the time-invariance 
of the false vacuum state is broken by a time-dependent field.
A first-order phase transition would produce a very inhomogeneous Universe\cite{Guth:1982pn},
unless a time-dependent scalar field leads to a rapidly changing percolation rate\cite{La:1989za,Linde:1990gz,Adams:1990ds}.
However, a second-order phase transition\cite{Linde:1981mu,Albrecht:1982wi}, 
controlled by a slowly-rolling scalar field, can lead to a smooth classical 
exit from the vacuum-dominated phase.

As a spectacular bonus, quantum fluctuations in that scalar field could provide 
a source of almost scale-invariant density 
fluctuations\cite{Press:1980zz,*Hawking:1982cz,*Starobinsky:1982ee,*Guth:1982ec,Bardeen:1983qw}, 
as detected in the CMB (see Chap. \ref{microwave}), which are thought to be the origin of the 
structures seen in the Universe today.

Accelerated expansion and primordial perturbations can also be produced in some modified 
gravity theories ({\textit{e.g.}},\cite{Starobinsky:1980te,Mukhanov:1981xt}), which introduce 
additional non-minimally coupled degrees of freedom.
Such inflation models can often be conveniently studied by transforming variables to 
an `Einstein frame' in which Einstein's equations apply with minimally coupled scalar 
fields\cite{Stelle:1977ry,Whitt:1984pd,Wands:1993uu}.

In the following we will review scalar field cosmology in general relativity and the 
spectra of primordial fluctuations produced during inflation, before studying selected 
inflation models.

\section{Scalar Field Cosmology}
\index{Cosmology!scalar field}%
\index{Scalar field cosmology}%

The energy-momentum tensor for a canonical scalar field $\phi$ with self-interaction 
potential $V(\phi)$ is given in \Eq{bigbang:eq:emomphi} in the ``Big-Bang Cosmology'' 
review.
In a homogeneous background this corresponds to a perfect fluid with density
\begin{equation}\label{inflation:eq:eq4}
\rho = \frac12 \dot\phi^2 + V(\phi) \,,
\end{equation}
and isotropic pressure
\begin{equation}\label{inflation:eq:eq5}
p = \frac12 \dot\phi^2 - V(\phi) \,,
\end{equation}
while the 4-velocity is proportional to the gradient of the field, $u^\mu\propto \nabla^\mu\phi$.

A field with vanishing potential energy acts like a stiff fluid with
$p=\rho=\dot\phi^2/2$, whereas if the time-dependence vanishes we have $p=-\rho=-V$ 
and the scalar field is uniform in time and space. Thus a classical, potential-dominated 
scalar-field cosmology, with $p\simeq -\rho$, can naturally drive a quasi-de Sitter 
expansion; the slow time-evolution of the energy density weakly breaks the exact 
$O(1,3)$ symmetry of four-dimensional de Sitter spacetime down to a Robertson-Walker 
(RW) spacetime, where the scalar field plays the role of the cosmic time coordinate.


In a scalar-field RW cosmology
the Friedmann constraint equation (\Eq{inflation:eq:FriedmannInf}) reduces to
\begin{equation}\label{inflation:eq:Hubblephi}
H^2 = \frac{8\pi}{3M_{\rm P}^2}  \left( \frac12 \dot\phi^2 + V \right) - \frac{k}{R^2} \,,
\end{equation}
while energy conservation (\Eq{bigbang:eq:econ} in ``Big-Bang Cosmology'' review) 
for a homogeneous scalar field reduces to the Klein-Gordon equation of motion 
(\Eq{bigbang:eq:phidot} in ``Big-Bang Cosmology'' review)
\begin{equation}\label{inflation:eq:KGeqn}
\ddot\phi = - 3H\dot\phi - V'(\phi) \,.
\end{equation}
The evolution of the scalar field is thus driven by the potential gradient 
$V'=dV/d\phi$, subject to damping by the Hubble expansion
$3H\dot\phi$.

If we define the Hubble slow-roll parameter
\begin{equation}\label{inflation:eq:epsilonH}
\epsilon_H \equiv - \frac{\dot{H}}{H^2}
\,,
\end{equation}
then we see that inflation ($\ddot{R}>0$ and hence $\dot{H}>-H^2$) requires $\epsilon_H<1$.
In this case the spatial curvature decreases relative to the scalar field energy density as the Universe expands.
Hence, in the following we drop the spatial curvature and consider a spatially-flat 
RW cosmology, assuming that inflation has lasted sufficiently long that our observable
Universe is very close to spatial flatness. However, we
note that bubble nucleation, leading to a first-order phase transition during 
inflation, can lead to homogeneous hypersurfaces with a hyperbolic (`open') 
geometry, effectively resetting the spatial curvature inside the bubble\cite{Coleman:1980aw}. 
This is the basis of so-called open inflation models\cite{Sasaki:1993ha,Bucher:1994gb,Linde:1995rv}, 
where inflation inside the bubble has a finite duration, leaving a finite negative spatial curvature.

In a scalar field-dominated cosmology, \Eq{inflation:eq:epsilonH} gives
\begin{equation}\label{inflation:eq:eq6}
\epsilon_H  = \frac{3\dot\phi^2}{2V+\dot\phi^2} \,,
\end{equation}
in which case we see that inflation requires a potential-dominated expansion, $\dot\phi^2<V$.

\subsection{Slow-Roll Inflation}

\index{Inflation!slow-roll}%

It is commonly assumed that the field acceleration term, $\ddot\phi$, in 
\Eq{inflation:eq:KGeqn} can be neglected, in which case
one can give an approximate solution for the inflationary attractor\cite{Liddle:1994dx}. 
This slow-roll approximation reduces the second-order Klein-Gordon equation~\eqref{inflation:eq:KGeqn} to a first-order system, which is over-damped, with 
the potential gradient being approximately balanced against the Hubble damping:
\begin{equation}\label{inflation:eq:eq7}
3H\dot\phi \simeq -V' \,,
\end{equation}
and at the same time the Hubble expansion (\Eq{inflation:eq:Hubblephi}) is 
dominated by the potential energy
\begin{equation}\label{imflation:eq:eq8}
H^2 \simeq \frac{8\pi}{3M_{\rm P}^2}  V(\phi) \,,
\end{equation}
corresponding to $\epsilon_H\ll1$.

A necessary condition for the validity of the slow-roll approximation is that the 
potential slow-roll parameters
\begin{equation}\label{imflation:eq:SRparams}
\epsilon \equiv \frac{M_{\rm P}^2}{16\pi} \left( \frac{V'}{V} \right)^2  \,,\quad
\eta \equiv \frac{M_{\rm P}^2}{8\pi} \left( \frac{V''}{V} \right) \,,
\end{equation}
are small, {\textit{i.e.}}, $\epsilon\ll1$ and $|\eta|\ll1$, requiring the potential to be 
correspondingly flat. If we identify $V''$ with the effective mass of the field, 
we see that the slow-roll approximation requires that the mass of the scalar field 
must be small compared with the Hubble scale.
We note that the Hubble slow-roll parameter 
(\Eq{inflation:eq:epsilonH}) 
coincides with the potential slow-roll parameter, $\epsilon_H\simeq\epsilon$, to 
leading order in the slow-roll approximation.

The slow-roll approximation allows one to determine the Hubble expansion rate as 
a function of the scalar field value, and vice versa. In particular, we can express, 
in terms of the scalar field value during inflation, the total logarithmic expansion, 
or number of ``e-folds'':
\begin{align}\label{imflation:eq:eq9}
N_* &\equiv \ln \left( \frac{R_{\rm end}}{R_*} \right) \cr
&= \int_{t_*}^{t_{\rm end}} H dt \simeq - \int_{\phi_*}^{\phi_{\rm end}} \sqrt{\frac{4\pi}{\epsilon}} \frac{d\phi}{M_{\rm P}} \ {\textrm{for}}\ V'>0
\,.\cr
\end{align}
Given that the slow-roll parameters are approximately constant during slow-roll inflation, 
$d\epsilon/dN\simeq2\epsilon(\eta-2\epsilon)={\cal O}(\epsilon^2)$, we have
\begin{equation}\label{imflation:eq:eq10}
N_* \simeq \frac{4}{\sqrt\epsilon} \frac{\Delta\phi}{M_{\rm P}} \, .
\end{equation}
Since we require $N>40$ to solve the flatness, horizon, and entropy problems of the 
standard Big-Bang cosmology, we require either very slow roll,  $\epsilon< 0.01$, 
or a large change in the value of the scalar field relative to the Planck scale, 
$\Delta\phi> M_{\rm P}$.

\subsection{Reheating}

\index{Inflation!reheating}%

Slow-roll inflation can lead to an exponentially large Universe, close to spatial 
flatness and homogeneity, but the energy density is locked in the potential energy 
of the scalar field, and needs to be converted to particles and thermalised to 
recover a hot Big-Bang cosmology at the end of inflation\cite{Kofman:1997yn,Bassett:2005xm}. 
This process is usually referred to as reheating, although there was not necessarily any 
preceding thermal era. Reheating can occur when the scalar field evolves towards the 
minimum of its potential, converting the potential energy first to kinetic energy. 
This can occur either through the breakdown of the slow-roll condition in single-field 
models, or due to an instability triggered by the inflaton reaching a critical value, 
in multi-field models known as hybrid inflation~\cite{Linde:1993cn}.
\index{Hybrid inflation model}%

Close to a simple minimum, the scalar field potential can be described by a quadratic 
function, $V=m^2\phi^2/2$, where $m$ is the mass of the field.
We can obtain slow-roll inflation in such a potential at large field values, 
$\phi\gg M_{\rm P}$. However,
for $\phi\ll M_{\rm P}$ the field approaches an oscillatory solution:
\begin{equation}\label{imflation:eq:eq11}
\phi(t) \simeq \frac{M_{\rm P}}{\sqrt{3\pi}} \frac{\sin(mt)}{mt} \,.
\end{equation}
For $|\phi|< M_{\rm P}$ the Hubble rate drops below the inflaton mass, $H<m$, and the 
field oscillates many times over a Hubble time. Averaging over several oscillations, 
$\Delta t\gg m^{-1}$, we find $\langle\dot\phi^2/2\rangle_{\Delta t}\simeq\langle m^2\phi^2/2\rangle_{\Delta t}$ and hence
\begin{equation}\label{imflation:eq:eq12}
\langle\rho\rangle_{\Delta t} \simeq \frac{M_{\rm P}^2}{6\pi t^2} \,,\quad
\langle p\rangle_{\Delta t} \simeq 0 \,.
\end{equation}
This coherent oscillating field corresponds to a condensate of non-relativistic 
massive inflaton particles, driving a matter-dominated era at the end of inflation, 
with scale factor $R\propto t^{2/3}$.

\index{Inflaton scalar field}%
The inflaton condensate can lose energy through perturbative decays due to terms 
in the interaction Lagrangian, such as
\begin{equation}\label{imflation:eq:eq13}
{\cal L}_{\rm int} \subset - \lambda_i \sigma\phi\chi_i^2 - \lambda_j\phi\bar\psi_j\psi_j
\end{equation}
that couple the inflation to scalar fields $\chi_i$ or fermions $\psi_j$, where 
$\sigma$ has dimensions of mass and the $\lambda_i$ are dimensionless couplings. 
When the mass of the inflaton is much larger than the decay products, the decay 
rate is given by\cite{Dolgov:1982th}
\begin{equation}\label{imflation:eq:eq14}
\Gamma_i = \frac{\lambda_i^2\sigma^2}{8\pi m} \, ,\quad
\Gamma_j = \frac{\lambda_j^2 m}{8\pi} \, .\
\end{equation}
These decay products must in turn thermalise with Standard Model particles before 
we recover conventional hot Big-Bang cosmology. An upper limit on the reheating 
temperature after inflation is given by\cite{Bassett:2005xm}
\begin{equation}\label{imflation:eq:eq15}
T_{\rm rh} = 0.2 \left( \frac{100}{g_*} \right)^{1/4} \sqrt{M_{\rm P}\Gamma_{\rm tot}} \, ,
\end{equation}
where $g_*$ is the effective number of degrees of freedom and $\Gamma_{\rm tot}$ 
is the total decay rate for the inflaton, which is required to be less than $m$ 
for perturbative decay.

The baryon asymmetry of the Universe must be generated after the main release of 
entropy during inflation, which is an important constraint on possible models.
Also, the fact that the inflaton mass is much larger than the mass scale of the 
Standard Model opens up the possibility that it may decay into massive stable or 
metastable particles that could be connected with dark matter, constraining 
possible models.
For example, in the context of supergravity models the reheat temperature is 
constrained by the requirement that gravitinos are not overproduced, potentially 
destroying the successes of Big-Bang nucleosynthesis. For a range of gravitino 
masses one must require $T_{\rm rh}<10^9$~GeV~\cite{Ellis:1983ew,Kawasaki:1994af}.

The process of inflaton decay and reheating can be significantly altered by 
interactions leading to space-time dependences in
the effective masses of the  fields. In particular, parametric resonance can 
lead to explosive, non-perturbative decay of the inflaton in some cases, a 
process often referred to as preheating\cite{Traschen:1990sw,Kofman:1997yn}. 
For example, an interaction term of the form
\begin{equation}\label{inflation:eq:eq16}        
{\cal L}_{\rm int} \subset - \lambda^2 \phi^2 \chi^2 \,,
\end{equation}
leads to a time-dependent effective mass for the $\chi$ field as the inflaton 
$\phi$ oscillates. This can lead to non-adiabatic particle production if the 
bare mass of the $\chi$ field is small for large couplings or for rapid changes 
of the inflaton field. The process of preheating is highly model-dependent, 
but it highlights the possible role of non-thermal particle production after 
and even during inflation.

\section{Primordial Perturbations from Inflation}
\label{inflation:sec:primodialPert}

\index{Inflation primordial perturbations}%

Although inflation was originally discussed as a solution to the problem of 
initial conditions required for homogeneous and isotropic hot Big-Bang 
cosmology, it was soon realised that inflation also offered a mechanism to 
generate the inhomogeneous initial conditions required for the formation of 
large-scale structure~\cite{Press:1980zz,*Hawking:1982cz,*Starobinsky:1982ee,*Guth:1982ec,Bardeen:1983qw,Mukhanov:1981xt,Chibisov:1982nx}.

\subsection{Metric Perturbations}

\index{Inflation primordial perturbations!metric}%
\index{Inflation!quantum fluctuations}%
\index{Quantum!fluctuations, inflation}%
In a homogeneous classical inflationary cosmology driven by a scalar field, 
the inflaton field is uniform on constant-time hypersurfaces, $\phi=\phi_0(t)$. 
However, quantum fluctuations inevitably break the spatial symmetry leading 
to an inhomogeneous field:
\begin{equation}\label{inflation:eq:eq17}
\phi(t,x^i)=\phi_0(t)+\delta\phi(t,x^i) \,.
\end{equation}
At the same time, one should consider inhomogeneous perturbations of the RW 
spacetime metric (see, {\textit{e.g.}}, Refs.~\cite{Kodama:1985bj,inflation:Mukhanov:1990me,Malik:2008im}):
\begin{equation}\label{inflation:eq:pertmetric}
\begin{split}
ds^2 = (1+2A)dt^2 & - 2R B_i dtdx^i \cr& - R^2 \left[ (1+2C)\delta_{ij} + \partial_i\partial_j E + h_{ij} \right]
dx^idx^j,\cr
\end{split}
\end{equation}
where $A$, $B$, $C$ and $E$ are scalar perturbations while $h_{ij}$ represents 
transverse and tracefree, tensor metric perturbations. Vector metric perturbations 
can be eliminated using Einstein constraint equations in a scalar field cosmology.

The tensor perturbations remain invariant under a temporal gauge transformation 
$t\to t+\delta t(t,x^i)$, but both the scalar field and the scalar metric 
perturbations transform. For example, we have
\begin{equation}\label{inflation:eq:eq18}
\delta\phi \to \delta\phi - \dot\phi_0 \delta t \,,\quad
C \to C - H \delta t \,.
\end{equation}
However, there are gauge invariant combinations, such as\cite{Mukhanov:1988jd}
\begin{equation}\label{inflation:eq:eq19}
Q = \delta\phi -\frac{\dot\phi_0}{H} C \,,
\end{equation}
which describes the scalar field perturbations on spatially-flat
($C=0$)
hypersurfaces. This is simply related to the curvature perturbation on uniform-field
($\delta\phi=0$)
hypersurfaces:
%
\begin{equation}\label{inflation:eq:Rcomoving}
{\cal R} = C - \frac{H}{\dot\phi_0} \delta\phi = - \frac{H}{\dot\phi_0} Q \,,
\end{equation}
which coincides in slow-roll inflation, $\rho \simeq \rho(\phi)$, with the 
curvature perturbation on uniform-density hypersurfaces\cite{Bardeen:1983qw}
\begin{equation}\label{inflation:eq:zeta}
\zeta = C - \frac{H}{\dot\rho_0} \delta\rho \,.
\end{equation}
Thus scalar field and scalar metric perturbations are coupled by the evolution 
of the inflaton field.

\subsection{Gravitational waves from inflation}

\index{Inflation primordial perturbations!gravitational waves}%

The tensor metric perturbation, $h_{ij}$ in
\Eq{inflation:eq:pertmetric}, is gauge-invariant and decoupled from the scalar 
perturbations at first order.  This represents the free excitations of the 
spacetime, {\textit{i.e.}}, gravitational waves, which are the simplest metric perturbations 
to study at linear order.

Each tensor mode, with wavevector $\vec{k}$, has two linearly-independent 
transverse and trace-free polarization states:
\begin{equation}\label{inflation:eq:eq20}
h_{ij} (\vec{k}) = h_{\vec{k}} q_{ij} + \bar{h}_{\vec{k}} \bar{q}_{ij} \,.
\end{equation}
The linearised Einstein equations then yield the same evolution equation for the 
amplitude as that for a massless field in RW spacetime:
\begin{equation}\label{inflation:eq:eq21}
\ddot{h}_{\vec{k}} + 3H\dot{h}_{\vec{k}}+ \frac{k^2}{R^2} h_{\vec{k}} = 0 \,,
\end{equation}
(and similarly for $ \bar{h}_{\vec{k}}$).
This can be re-written in terms of the conformal time, $\eta=\int dt/R$, and the 
conformally rescaled field:
\begin{equation}\label{inflation:eq:eq22}
u_{\vec{k}} = \frac{M_{\rm P} R h_{\vec{k}}}{\sqrt{32\pi}} \,.
\end{equation}
This conformal field then obeys the wave equation for a canonical scalar field in 
Minkowski spacetime with a time-dependent mass:
\begin{equation}\label{inflation:eq:scalar-pert-eom}
u_{\vec{k}}'' + \left( k^2 - \frac{R''}{R} \right) u_{\vec{k}} = 0 \,.
\end{equation}
During slow roll
\begin{equation}\label{inflation:eq:eq23}
\frac{R''}{R} \simeq (2-\epsilon) R^2H^2 \,.
\end{equation}
This makes it possible
to quantise the linearised metric fluctuations, $u_{\vec{k}}\to\hat{u}_{\vec{k}}$, 
on sub-Hubble scales, $k^2/R^2 \gg H^2$, where the background expansion can be neglected.

Crucially, in an inflationary expansion, where $\ddot{R}>0$, the comoving Hubble 
length $H^{-1}/R=1/\dot{R}$ decreases with time. Thus, all modes start inside the 
Hubble horizon and it is possible to take the initial field fluctuations to be in 
a vacuum state at early times or on small scales:
\begin{equation}\label{inflation:eq:eq24}
\langle u_{\vec{k}_1} u_{\vec{k}_2} \rangle = \frac{i}{2} (2\pi)^3 \delta^{(3)}\left(\vec{k}_1+\vec{k}_2\right) \,.
\end{equation}

\pdgwidefigure{Planck18nsr.pdf}{
The marginalized joint 68 and 95\% CL regions for the tilt in the scalar 
perturbation spectrum, $n_{\rm s}$, and the relative magnitude of the tensor 
perturbations, $r$, obtained from the \textit{Planck} 2018 and lensing data 
alone, and their combinations with BICEP2/Keck Array (BK15) and (optionally) 
BAO data, confronted with the predictions of some of the inflationary models 
discussed in this review. This figure is taken from~Ref.\protect\cite{Akrami:2018odb}.}
{inflation:fig:Planck2018}{}{width=0.95\linewidth}

In terms of the amplitude of the tensor metric perturbations, this corresponds to
\begin{equation}\label{inflation:eq:eq25}
\langle h_{\vec{k}_1} h_{\vec{k}_2} \rangle = \frac12 
\frac{{\cal P}_{\rm t}(k_1)}{4\pi k_1^3} (2\pi)^3 \delta^{(3)} \left( \vec{k}_1+\vec{k}_2 \right) \,,
\end{equation}
where the factor $1/2$ appears due to the two polarization states that contribute 
to the total tensor power spectrum:
\begin{equation}\label{inflation:eq:eq26}
{\cal P}_{\rm t} (k) = \frac{64\pi}{M_{\rm P}^2} \left( \frac{k}{2\pi R} \right)^2 \,.
\end{equation}
%

On super-Hubble scales, $k^2/R^2\ll H^2$, we have the growing mode solution to 
\Eq{inflation:eq:scalar-pert-eom}, $u_{\vec{k}}\propto R$, corresponding to 
$h_{\vec{k}}\to$ constant, {\textit{i.e.}}, tensor modes are frozen-in on super-Hubble scales, 
both during and after inflation. Thus, connecting the initial vacuum fluctuations 
on sub-Hubble scales to the late-time power spectrum for tensor modes at Hubble 
exit during inflation, $k=R_*H_*$, we obtain
\begin{equation}\label{inflation:eq:defPT}
{\cal P}_{\rm t} (k) \simeq \frac{64\pi}{M_{\rm P}^2} \left( \frac{H_*}{2\pi} \right)^2 \,.
\end{equation}

\index{Inflation!de Sitter limit $\epsilon\to0$}%

In the de Sitter limit, $\epsilon\to0$, the Hubble rate becomes time-independent 
and the tensor spectrum on super-Hubble scales becomes 
scale-invariant\cite{Starobinsky:1979ty}. However slow-roll evolution leads to 
weak time dependence of $H_*$ and thus a scale-dependent spectrum on large scales, 
with a spectral tilt
\begin{equation}\label{inflation:eq:nT}
n_{\rm t} \equiv \frac{d\ln {\cal P}_{\rm t}}{d\ln k} \simeq -2\epsilon_* \,.
\end{equation}

\subsection{Density Perturbations from single-field inflation}

\index{Inflation primordial perturbations!density from single-field}%
\index{Inflaton!field fluctuations}%
The inflaton field fluctuations on spatially-flat hypersurfaces are coupled to 
scalar metric perturbations at first order, but these can be eliminated using the 
Einstein constraint equations to yield an evolution equation
\begin{equation}\label{inflation:eq:eq27}
\ddot{Q}_{\vec{k}} + 3H\dot{Q}_{\vec{k}} + \left[ \frac{k^2}{R^2} + V'' - 
\frac{8\pi}{M_{\rm P}^2 R^3} 
\frac{d}{dt}\left( \frac{R^3\dot\phi^2}{H} \right) \right] Q_{\vec{k}} = 0 \,.
\end{equation}
Terms proportional to $M_{\rm P}^{-2}$ represent the effect on the field fluctuations 
of gravity at first order. As can be seen, this vanishes in the limit of a 
constant background field, and hence is suppressed in the slow-roll limit, but 
it is of the same order as the effective mass, $V''=3\eta H^2$, so must be 
included if we wish to model deviations from exact de Sitter symmetry.

This wave equation can also be written in the canonical form for a free field 
in Minkowski spacetime if we define\cite{Mukhanov:1988jd}
\begin{equation}\label{inflation:eq:eq28}
v_{\vec{k}} \equiv R Q_{\vec{k}} \,,
\end{equation}
to yield
\begin{equation}\label{inflation:eq:eq29}
v_{\vec{k}}'' + \left( k^2 - \frac{z''}{z} \right) v_{\vec{k}} = 0 \,,
\end{equation}
where we further define
\begin{equation}\label{inflation:eq:eq30}
z \equiv \frac{R\dot\phi}{H} \,, \qquad \frac{z''}{z} \simeq (2+5\epsilon-3\eta) R^2 H^2 \, ,
\end{equation}
with the last approximate equality holding to leading order in the slow-roll approximation.

As previously done for gravitational waves, we quantise the linearised field 
fluctuations $v_{\vec{k}}\to\hat{v}_{\vec{k}}$ on sub-Hubble scales, $k^2/R^2 \gg H^2$, 
where the background expansion can be neglected. Thus, we impose
%
\begin{equation}\label{inflation:eq:eq31}
\langle v_{\vec{k}_1} v_{\vec{k}_2}' \rangle = \frac{i}{2} \delta^{(3)}\left(\vec{k}_1+\vec{k}_2\right) \,.
\end{equation}
In terms of the field perturbations, this corresponds to
\begin{equation}\label{inflation:eq:eq32}
\langle Q_{\vec{k}_1} Q_{\vec{k}_2} \rangle = \frac{{\cal P}_Q(k_1)}{4\pi k_1^3} (2\pi)^3 \delta^{(3)} \left( \vec{k}_1+\vec{k}_2 \right) \,,
\end{equation}
where the power spectrum for vacuum field fluctuations on sub-Hubble scales, 
$k^2/R^2 \gg H^2$, is simply
\begin{equation}\label{inflation:eq:eq33}
{\cal P}_Q (k) =  \left( \frac{k}{2\pi R} \right)^2 \,,
\end{equation}
yielding the classic result for the vacuum fluctuations for a massless field in 
de Sitter at Hubble exit, $k=R_*H_*$:
\begin{equation}\label{inflation:eq:eq34}
{\cal P}_Q (k) \simeq  \left( \frac{H}{2\pi} \right)_*^2 \, .
\end{equation}
In practice there are slow-roll corrections due to the small but finite mass ($\eta$) 
and field evolution ($\epsilon$)\cite{Stewart:1993bc}.

Slow-roll corrections to the field fluctuations are small on sub-Hubble scales, but can 
become significant as the field and its perturbations evolve over time on super-Hubble 
scales. This is why it is helpful to work instead with the curvature perturbation, $\zeta$ 
defined in \Eq{inflation:eq:zeta}, which remains constant on super-Hubble 
scales for adiabatic density perturbations both during and after 
inflation\cite{Bardeen:1983qw,Wands:2000dp}. 
Thus we have an expression for the primordial curvature perturbation on super-Hubble 
scales produced by single-field 
slow-roll inflation:
\begin{equation}\label{inflation:eq:defPzeta}
{\cal P}_\zeta (k) = \left[ \left( \frac{H}{\dot\phi} \right)^2 {\cal P}_Q (k) \right]_* \simeq  
\frac{4\pi}{M_{\rm P}^2} \left[ \frac{1}{\epsilon} \left( \frac{H}{2\pi} \right)^2 \right]_* \, .
\end{equation}
Comparing this with the primordial gravitational wave power spectrum (\Eq{inflation:eq:defPT}) 
we obtain the tensor-to-scalar ratio 
%
\begin{equation}\label{inflation:eq:singlefieldr}
r \equiv \frac{{\cal P}_{\rm t}}{{\cal P}_\zeta} \simeq 16\epsilon_* \,.
\end{equation}
Note that the scalar amplitude is boosted by a factor $1/\epsilon_*$ during slow-roll inflation, 
because small scalar field fluctuations can lead to relatively large curvature perturbations on 
hypersurfaces defined with respect to the density if the potential energy is only weakly 
dependent on the scalar field, as in slow roll. Indeed, the de Sitter limit is singular, 
since the potential energy becomes independent of the scalar field at first order, 
$\epsilon\to0$, and the curvature perturbation on uniform-density hypersurfaces 
becomes ill-defined.

We note that in single-field inflation the tensor-to-scalar ratio and the tensor 
tilt (\Eq{inflation:eq:nT}) at the same scale are both determined by the first 
slow-roll parameter at Hubble exit, $\epsilon_*$, giving rise to an important 
consistency test for single-field 
slow-roll inflation:
\begin{equation}\label{inflation:eq:eq35}
n_{\rm t} \simeq -\frac{r}{8} \,.
\end{equation}
This may be hard to verify if $r$ is small, making any tensor tilt $n_{\rm t}$ difficult 
to measure. On the other hand, it does offer a way to rule out single-field slow-roll 
inflation if either $r$ or $n_{\rm t}$ is large.

Given the relatively large scalar power spectrum, it has proved easier to measure the 
scalar tilt, conventionally defined as $n_{\rm s}-1$.
Slow-roll corrections lead to slow time-dependence of both $H_*$ and $\epsilon_*$, 
giving a weak scale-dependence of the scalar power spectrum:
\begin{equation}\label{inflation:eq:ns}
n_{\rm s}  - 1 \equiv \frac{d\ln {\cal P}_\zeta}{d\ln k} \simeq -6\epsilon_* +2\eta_* \,,
\end{equation}
and a running of this tilt at second-order in slow roll:
\begin{equation}\label{inflation:eq:eq36}
\frac{d n_{\rm s}}{d\ln k} \simeq -8\epsilon_* (3\epsilon_*-2\eta_*) -2\xi_*^2 \,,
\end{equation}
where the running introduces a new slow-roll parameter at second order: 
\begin{equation}\label{inflation:eq:eq37}
\xi^2 = \frac{M_{\rm P}^4}{64\pi^2} \frac{V'V'''}{V^2} \,.
\end{equation}

\subsection{Observational Bounds}

\index{Inflation primordial perturbations!observational bounds}%

The observed scale-dependence of the power spectrum makes it necessary to specify 
the comoving scale, $k$, at which quantities are constrained and hence the Hubble-exit 
time, $k=a_*H_*$, when the corresponding theoretical quantities are calculated during 
inflation. This is usually expressed in terms of the number of e-folds from the end 
of inflation\cite{Liddle:2003as}:
\begin{align}\label{inflation:eq:nefolds}
N_*(k) \simeq 67 &- \ln \left( \frac{k}{a_0H_0} \right) + \frac14 \ln \left( \frac{V_*^2}{M_{\rm P}^4\rho_{\rm end}} \right) 
+ \frac{1}{12} \ln \left( \frac{\rho_{\rm rh}}{\rho_{\rm end}} \right) \cr&- \frac{1}{12}\ln(g_*),
\end{align}
where $H_0^{-1}/a_0$ is the present comoving Hubble length.
Different models of reheating and thus different reheat temperatures and densities, 
$\rho_{\rm rh}$ in ~\Eq{inflation:eq:nefolds}, lead to a range of possible values for $N_*$ 
corresponding to a fixed physical scale, and hence we have a range of observational 
predictions for a given inflation model;
this is illustrated by allowing a fixed range of values for $N_*$ in~\Fig{inflation:fig:Planck2018}.


\pdgwidefigure{reconstruction.pdf}{
The result of reconstructing a single-field inflaton potential using a cubic-spline 
power-spectrum 
mode expansion and the full \textit{Planck}, lensing, BK15
and BAO data set. This figure is taken
from Refs.~\protect\cite{Akrami:2018odb}.}
{inflation:fig:reconstruction}{!htb}{width=0.9\linewidth}

The \textit{Planck} 2018 temperature and polarization data (see Chap. \ref{microwave}, 
``Cosmic Microwave Background'' review)
are consistent with a smooth featureless power spectrum over a range of comoving 
wavenumbers, $0.008\ h^{-1}$~Mpc$^{-1} \leq k\leq 0.1\ h$~Mpc$^{-1}$. In the absence 
of running, the {\textit{Planck}} data measure the spectral index to be~\cite{Akrami:2018odb}
\begin{equation}\label{inflation:eq:eq38}
n_{\rm s} = 0.9649 \pm 0.0042 \,,
\end{equation}
corresponding to a deviation from scale-invariance exceeding the $7\sigma$ level. 
Recent results from the Atacama Cosmology Telescope (ACT)~\cite{ACT:2025fju} on small angular scales in combination with {\textit{Planck}} data on larger scales, lensing measurements and DESI DR2 data~\cite{DESI:2025zgx} suggest a somewhat higher value, $n_{\rm s} = 0.975 \pm 0.003$, although this is not yet a full joint analysis. There are also recent results from the South Pole Telescope~\cite{SPT-3G:2025bzu} that are quite consistent with {\textit{Planck}}.
If running of the spectral tilt is included in the model, this is constrained to 
be~\cite{Akrami:2018odb}
\begin{equation}\label{inflation:eq:e39}
\frac{d n_{\rm s}}{d\ln k} = - 0.0045 \pm 0.0067
\end{equation}
at the 95\% CL, assuming no running of the running.
The most recent analysis~\cite{BICEPKeck:2021gln} of the BICEP, Keck Array and {\textit{Planck}} data places an upper bound on the tensor-to-scalar ratio
\begin{equation}\label{inflation:eq:eq40}
r < 0.036 \,,
\end{equation}
at the 95\% CL.

These observational bounds can be converted into bounds on the slow-roll parameters 
and hence the potential during slow-roll inflation. Setting higher-order slow-roll 
parameters (beyond second-order in horizon-flow parameters~\cite{inflation:leach/etal:2002}) 
to zero, the \textit{Planck} collaboration 
obtain the following 95\% CL bounds when 
lensing and BK15 data are included~\cite{Akrami:2018odb}
\begin{align}                                                            
\epsilon &< 0.0044 \,, \label{inflation:eq:eq41}\\
\eta &= -0.015 \pm 0.006 \,, \label{inflation:eq:eq42}\\                                                                            
\xi^2 &= 0.0029^{+0.0073}_{-0.0069} \,, \label{inflation:eq:eq43} 
\end{align}
which can be used to constrain models, as discussed in the next section.

Figure \ref{inflation:fig:Planck2018},
which is taken from Ref.~\cite{Akrami:2018odb}, compares
observational CMB constraints on the tilt, $n_{\rm s}$, in the spectrum of scalar perturbations
and the ratio, $r$, between the magnitudes of tensor and scalar perturbations.
Important r\^oles are played by data from the 
\index{CMB!{\it Planck} satellite}%
\index{Satellite!{\it Planck}, CMB}%
\textit{Planck} satellite and on lensing, the BICEP2/Keck Array (BK15)
and measurements of 
\index{Baryon!acoustic oscillations}%
baryon acoustic oscillations (BAO). The reader is referred to Ref.~\cite{Akrami:2018odb}
for technical details. These experimental constraints are compared
with the predictions of some of the inflationary models discussed in this review. Generally speaking,
models with a concave potential are favored over those with a convex potential, and
models with power-law inflation are now excluded, as opposed to models with de Sitter-like (quasi-)exponential
expansion.

There is no significant evidence for local features within the range of inflaton field values probed by the data~\cite{Akrami:2018odb}.
However, the data may be used to partially reconstruct the effective inflationary potential over
a range of inflaton field values, assuming that it is suitably smooth. 
The result of one such exercise by the \textit{Planck} collaboration~\cite{Akrami:2018odb} 
in the framework of a generic single-field inflaton potential is shown in \Fig{inflation:fig:reconstruction}.
This reconstruction assumes a cubic-spline power-spectrum mode expansion and employs the full \textit{Planck}, lensing, BK15
and BAO data set. The reader is again referred to~\cite{Akrami:2018odb} for technical details. 
We see that the effective inflaton potential is relatively well reconstructed over field
values $\phi$ within $\pm 0.5$ of the chosen pivot value, but the potential is only very weakly constrained for
larger values of $|\phi - \phi_{\rm pivot}|$, providing wide scope for inflationary model-builders.

\section{Models}

\subsection{Pioneering Models}

\index{Inflation models!pioneering}%
The paradigm of the inflationary Universe was proposed in Ref.~\cite{Guth:1980zm},
where it was pointed out that an early period of (near-)exponential expansion, in addition to
resolving the horizon and flatness problems of conventional Big-Bang
cosmology as discussed above, would also dilute the prior abundance of any unseen heavy,
(meta-)stable particles, as exemplified by monopoles in grand unified
theories (GUTs; see Chap.~\ref{guts}, ``Grand Unified Theories'' review). 
The possibility of a de Sitter phase in the early history of the Universe was
also proposed in the non-minimal gravity model of Ref.~\cite{Starobinsky:1980te}, with the motivation of avoiding an
initial singularity.
The original proposal in Ref.~\cite{Guth:1980zm} was that this inflationary expansion took
place while the Universe was in a metastable state (a similar
suggestion was made in Refs.~\cite{Kazanas:1980tx,Sato:1980yn}, where in Ref.~\cite{Kazanas:1980tx} it was also pointed out that such a
mechanism could address the horizon problem) and was terminated by a first-order transition
due to tunneling though a potential barrier. However, it was recognized
already in Ref.~\cite{Guth:1980zm} that this `old inflation' scenario would need modification if the
transition to the post-inflationary Universe were to be completed smoothly without
generating unacceptable inhomogeneities.

This `graceful exit' problem was addressed in the `new inflation' model
of Ref.~\cite{Linde:1981mu} (see also Ref.~\cite{Albrecht:1982wi} and footnote [39] of Ref.~\cite{Guth:1980zm}),
which studied models based on an SU(5) GUT with an effective potential
of the Coleman-Weinberg type ({\textit{i.e.}}, dominated by radiative corrections),
in which inflation could occur during the roll-down from the local maximum
of the potential towards a global minimum. However, it was realized that the
Universe would evolve to a different minimum from the Standard Model\cite{Billoire:1981fw,*Breit:1983pp},
and it was also recognized that density fluctuations would necessarily be too large~\cite{Press:1980zz,*Hawking:1982cz,*Starobinsky:1982ee,*Guth:1982ec},
since they are related to the GUT coupling strength.

These early models of inflation assumed initial conditions enforced by
thermal equilibrium in the early Universe. However, this assumption was
questionable: indeed, it was not made in the model of Ref.~\cite{Starobinsky:1980te},
in which a higher-order gravitational curvature term was assumed to
arise from quantum corrections, and the assumption of initial thermal
equilibrium was jettisoned in the `chaotic' inflationary model of Ref.~\cite{Linde:1983gd}.
These have been the inspirations for much recent inflationary model building,
so we now discuss them in more detail, before reviewing contemporary
models.

In this section we will work in natural units where we set the reduced Planck mass 
to unity, {\textit{i.e.}}, $8\pi/M_{\rm P}^2=1$. All masses are thus relative to the reduced Planck scale.

\subsection{${\rm R}^2$ Inflation}
\label{inflation:models:R2}

\index{Inflation models!${\rm R}^2$ inflation}%

The first-order Einstein-Hilbert action,
$(1/2) \int d^4x \sqrt{-g} \R$,
where $\R$ is the Ricci scalar
curvature, is the minimal possible theory consistent with general coordinate invariance.
However, it is possible that there might be non-minimal corrections to this action, and the
unique second-order possibility is
\begin{equation}\label{inflation:eq:Starob}
S=\frac{1}{2} \int d^4x \sqrt{-g} \left( \R+\frac{\R^2}{6M^2} \right) \, .
\end{equation}
It was pointed out in Ref.~\cite{Starobinsky:1980te} that an $\R^2$ term could be generated by
quantum effects, and that \Eq{inflation:eq:Starob} could lead to de Sitter-like expansion of the Universe.
Scalar density perturbations in this model were calculated in Ref.~\cite{Mukhanov:1981xt}. Because the
initial phase was (almost) de Sitter, these perturbations were (approximately) scale-invariant, with
magnitude $\propto M$. It was also pointed out in Ref.~\cite{Mukhanov:1981xt} that requiring the
scalar density perturbations to lie in the range $10^{-3}$ to $10^{-5}$, consistent with
upper limits at that time, would require $M \sim 10^{-3}$ to $10^{-5}$ in Planck units, and it was further suggested
that these perturbations could lead to the observed large-scale structure
of the Universe, including the formation of galaxies.

Although the action~\eqref{inflation:eq:Starob} does not contain an explicit scalar field, Ref.~\cite{Mukhanov:1981xt}
reduced the calculation of density perturbations to that of fluctuations in the scalar
curvature $\R$, which could be identified (up to a factor) with a scalar field of mass $M$.
The formal equivalence of $\R^2$ gravity~\eqref{inflation:eq:Starob} to
a theory of gravity with a massive scalar $\phi$ had been shown in Ref.~\cite{Stelle:1977ry} (see also Ref.~\cite{Whitt:1984pd}).
The effective scalar potential for what we would nowadays call the `inflaton'~\cite{Nanopoulos:1983up}
takes the form
\begin{equation}\label{inflation:eq:Staropot}
S=\frac{1}{2} \int d^4x \sqrt{-{g}} \left[{\R} + (\partial_\mu \phi)^2 - \frac{3}{2} M^2 (1- e^{-\sqrt{2/3}\phi})^2 \right] \, 
\end{equation}
when the action is written in the Einstein frame, and the potential is shown as the solid black line in \Fig{inflation:fig:Staropot}.
Using \Eq{inflation:eq:defPzeta},
one finds that the amplitude of the scalar density perturbations in this model is given by
\begin{equation}\label{inflation:eq:StaroAs}
\Delta_{\cal R} =  \frac{3 M^2}{8\pi^2} \sinh^4 \left(\frac{\phi}{\sqrt{6}}\right) \, ,
\end{equation}
The measured magnitude of the density fluctuations
in the CMB requires $M \simeq 1.3 \times 10^{-5}$ in Planck units (assuming $N_* \simeq 55$), so
one of the open questions in this model is why $M$ is so small. Obtaining $N_* \simeq 55$
also requires an initial value of $\phi \simeq 5.5$, {\textit{i.e.}}, a super-Planckian initial condition,
and another issue for this and many other models is how the form of the effective potential
is protected and remains valid at such large field values. Using \Eq{inflation:eq:ns} one finds that
$n_{\rm s} \simeq 0.965$ for $N_* \simeq 55$ and using (\Eq{inflation:eq:singlefieldr}) one finds that $r \simeq 0.0035$.
These predictions are consistent with the present data from \textit{Planck} and other experiments, as seen in \Fig{inflation:fig:Planck2018}.


\pdgwidefigure{Staropot.pdf}
{Inflationary potential $V$ in the $\R^2$ model (solid black line) compared 
with its form in various no-scale models discussed in detail in Ref.~\protect\cite{Ellis:2013xoa} 
(dashed coloured lines).}
{inflation:fig:Staropot}{!htb}{width=0.9\linewidth}

\subsection{Chaotic Models with Power-Law Potentials}
\label{inflation:models:chaotic}

\index{Inflation models!chaotic with power-law potentials}%

As has already been mentioned, a key innovation in inflationary model-building
was the suggestion to abandon the questionable assumption of a thermal initial state,
and consider `chaotic' initial conditions with very general forms of potential~\cite{Linde:1983gd}. (Indeed,
the $\R^2$ model discussed above can be regarded as a prototype of this approach.) The
chaotic approach was first proposed in the context of a simple power-law potential of the form $\mu^{4 - \alpha} \phi^\alpha$,
and the specific example of $\lambda \phi^4$ was studied in Ref.~\cite{Linde:1983gd}. Such models make the
following predictions for the slow-roll parameters $\epsilon$ and $\eta$:
\begin{equation}\label{inflation:eq:chaotic}
\epsilon \; = \; \frac{1}{2}\left(\frac{\alpha}{\phi}\right)^2, \; \; \eta \; = \;
\frac{\alpha(\alpha-1)}{\phi^2} \, ,
\end{equation}
leading to the predictions
\begin{equation}\label{inflation:eq:eq44}
r \; \approx \; \frac{4 \alpha}{N_*} \ , \qquad n_{\rm s}-1 \; \approx \; - \frac{\alpha+2}{2N_*} \, ,
\end{equation}
which are shown in \Fig{inflation:fig:Planck2018} for some illustrative values of $\alpha$.
We note that the prediction of the original $\phi^4$ model lies out of the frame, with values of $r$ that
are too large and values of $n_{\rm s}$ that are too small. The $\phi^3$ model has similar problems,
and would in any case require modification in order to have a well-defined minimum. The simplest
possibility is $\phi^2$, but this is now also disfavored by the data, at the 95\% CL if only the \textit{Planck}
data are considered, and more strongly if other data are included, as seen in ~\Fig{inflation:fig:Planck2018}. (For
non-minimal models of quadratic inflation that avoid this problem, see, {\textit{e.g.}}, Ref.~\cite{Koivisto:2014gia}.)
Indeed, as can be seen in \Fig{inflation:fig:Planck2018}, all
single-field
models with a convex potential ({\textit{i.e.}}, one
curving upwards) are disfavored compared to models with a concave potential. 

\subsection{Hilltop Models}
\label{inflation:models:hilltop}

\index{Inflation models!Hilltop}%

This preference for a concave potential motivates interest in `hilltop' models~\cite{Boubekeur:2005zm}, whose starting-point is
a potential of the form
\begin{equation}\label{inflation:eq:hilltop}
V(\phi) \; = \; {M}^4 \left[ 1 - \left(\frac{\phi}{\mu}\right)^p + \dots \right] \, ,
\end{equation}
where the $\dots$ represent extra terms that yield a positive semi-definite potential.
To first order in the slow-roll parameters, when $x \equiv \phi/\mu$ is small, one has
\begin{equation}\label{inflation:eq:hilltop1}
n_{\rm s} \; \simeq \; 1 - p(p - 1)
\mu^{-2} 
\frac{x^{p-2}}{(1 - x^p)} - \frac{3}{8}r \, , \qquad
r \; \simeq \; 8p^2 
\mu^{-2}  
\frac{x^{2p-2}}{(1 - x^p)^2} \, .
\end{equation}
As seen in \Fig{inflation:fig:Planck2018}, a hilltop model with $p = 4$ can be
compatible with the \textit{Planck} and other measurements, if $\mu\gg M_{\rm P}$.

\subsection{D-Brane Inflation}
\label{inflation:models:brane}

\index{Inflation models!D-brane}%

Many scenarios for inflation involving extra dimensions have been proposed, {\textit{e.g.}}, the
possibility that observable physics resides on a three-dimensional brane, and that there is an
inflationary potential that depends on the distance between our brane and an antibrane, with a
potential of the form~\cite{Dvali:2001fw,*GarciaBellido:2001ky,*Kachru:2003sx}
\begin{equation}\label{inflation:eq:Dbrane}
V(\phi) \; = \; {M}^4 \left[ 1 - \left(\frac{\mu}{\phi}\right)^p + \dots \right] \, .
\end{equation}
In this scenario the effective potential vanishes in the limit $\phi \to \infty$, corresponding
to complete separation between our brane and the antibrane.
The predictions for $n_{\rm s}$ and $r$ in this model can be obtained from Eq.~(\ref{inflation:eq:hilltop1})
by exchanging $p \leftrightarrow - p$, and are also consistent with the \textit{Planck} and other data.

\subsection{Natural Inflation}
\label{inflation:models:natural}

\index{Inflation models!natural}%

Also seen in \Fig{inflation:fig:Planck2018} are the predictions of `natural inflation'~\cite{Adams:1992bn}, in which one postulates
a non-perturbative shift symmetry that suppresses quantum corrections, so that a hierarchically small scale of
inflation, $H \ll M_{\rm P}$, is technically natural. In the simplest models, there is a periodic potential
of the form
\begin{equation}\label{inflation:eq:natural}
V(\phi) \; = \; {M}^4 \left[ 1 + \cos \left( \frac{\phi}{f} \right) \right] \, ,
\end{equation}
where $f$ is a dimensional parameter reminiscent of an axion decay constant (see the next subsection)~\cite{Pajer:2013fsa},
which must have a value $> M_{\rm P}$. Natural inflation can yield predictions similar to quadratic inflation
(which are no longer favored, as already discussed), but can also yield an effective convex potential.
Thus, it may lead to values of $r$ that are acceptably small, but for values of $n_{\rm s}$ that are in tension with the data,
as seen in \Fig{inflation:fig:Planck2018}.

\subsection{Axion Monodromy Models}

\index{Inflation models!axion monodromy}%
The effective potentials in stringy models~\cite{Silverstein:2008sg,McAllister:2008hb} motivated by axion monodromy may be
of the form
\begin{equation}\label{inflation:eq:monodromy}
V(\phi) \; = \; \mu^{4 - \alpha} \phi^\alpha + {M}^4e^{- C \left( \frac{\phi}{\phi_0} \right)^{p_ {M}}}
\cos \left[ \gamma + \frac{\phi}{f} \left( \frac{\phi}{\phi_0} \right)^{p_f + 1} \right] \, ,
\end{equation}
where $\mu, {M}, f$ and $\phi_0$ are parameters with the dimension of mass,
and $C,\ p,\ p_{M},\ p_f$ and $\gamma$ are dimensionless constants,
generalizing the potential~\cite{Adams:1992bn} in the simplest models of natural inflation. The oscillations in
Eq.~(\ref{inflation:eq:monodromy}) are associated with the axion field, and powers $p_{M}, p_f \ne 0$
may arise from $\phi$-dependent evolutions of string moduli. Since the exponential prefactor
in Eq.~(\ref{inflation:eq:monodromy}) is due to non-perturbative effects that may be strongly suppressed,
the oscillations may be unobservably small. Specific string models having $\phi^\alpha$ with $\alpha = 4/3, 1$ or $2/3$
have been constructed in Refs.~\cite{Silverstein:2008sg,McAllister:2008hb}, providing some motivation for the low-power models
mentioned above.

As seen in \Fig{inflation:fig:Planck2018}, the simple axion monodromy models with the power $\alpha = 4/3$ or $1$ are
no longer compatible with the current CMB data at the 95\% CL, while $\alpha = 2/3$ is only marginally compatible at 95\% CL.
The \textit{Planck} Collaboration
has also searched for characteristic effects associated with the second term in Eq.~(\ref{inflation:eq:monodromy}),
such as a possible drift in the modulation amplitude (setting $p_{M} = C = 0$), and
a possible drifting frequency generated by $p_f \ne 0$, without finding any
compelling evidence~\cite{Akrami:2018odb}.

\subsection{Higgs Inflation}
\label{inflation:models:higgs}

\index{Inflation models!Higgs}%

Since the energy scale during inflation is commonly expected to lie between the Planck
and TeV scales, it may serve as a useful bridge with contacts both to string theory or
some other quantum theory of gravity, on the one side, and particle physics on the other side.
However, as the above discussion shows, much of the activity in building models of inflation has been
independent of specific connections with these subjects, though some examples of string-motivated models of inflation were mentioned above.

The most economical scenario for inflation might be to use as the inflaton the only established scalar
field, namely the Higgs field (see Chap.\ref{higgs}, ``Status of Higgs boson physics'' review). A specific model assuming a non-minimal coupling of the Higgs
field $h$ to gravity was constructed in~Ref.\cite{Bezrukov:2007ep}. Its starting point is the action
\begin{equation}\label{inflation:eq:HiggsInf}
S \; = \; \int d^4x \sqrt{-g} \left[ \frac{M^2 + \xi h^2}{2} \R + \frac{1}{2} \partial_\mu h \partial^\mu h - \frac{\lambda}{4} ( h^2 - v^2 )^2 \right] \, ,
\end{equation}
where $v$ is the Higgs vacuum expectation value. The model requires $\xi \gg 1$, in which case it can be rewritten in the Einstein frame as
\begin{equation}\label{inflation:eq:HiggsInf2}
S \; = \; \int d^4x \sqrt{-g} \left[ 
\frac{1}{2} \R 
+ \frac{1}{2} \partial_\mu \chi \partial^\mu \chi - U(\chi) \right] \, ,
\end{equation}
where the effective potential for the canonically-normalized inflaton field $\chi$ has the form
\begin{equation}\label{inflation:eq:Uchi}
U(\chi) \; = \; 
\frac{\lambda}{4 \xi^2} 
\left[ 1 + \exp \left( - 
\frac{2 \chi}{\sqrt{6} M_{\rm P}} 
\right) \right]^{-2} \, ,
\end{equation}
which is similar to the effective potential of the $\R^2$ model at large field values. As such, the model inflates
successfully if $\xi \simeq 5 \times 10^4 \ m_h/(\sqrt{2} v)$, where $m_h$ is the Higgs mass, with predictions for
$n_{\rm s}$ and $r$ that are indistinguishable from the predictions of the $\R^2$ model shown in \Fig{inflation:fig:Planck2018}.~\footnote{Although Higgs and $R^2$ inflation share the same inflaton potential, they may reheat the Universe through different mechanisms, resulting in different reheating temperatures and hence different values of $N_*$ that would lead to slightly different predictions; see the discussion around Eq.~\eqref{inflation:eq:nefolds}.}

This model is appealing because of its connection to particle physics, but must confront several issues. One is to understand the value of $\xi$, and another
is the possibility of unitarity violation. However, a more fundamental issue is whether the effective quartic
Higgs coupling is positive at the scale of the Higgs field during inflation. Extrapolations of the effective
potential in the Standard Model using the measured values of the masses of the Higgs boson and the top quark
indicate that probably $\lambda < 0$ at this scale~\cite{Buttazzo:2013uya}, though there are still significant uncertainties associated with the
appropriate input value of the top mass and the extrapolation to high renormalization scales.

\subsection{Supersymmetric Models of Inflation}

\index{Inflation models!supersymmetric}%

Supersymmetry~\cite{Nilles:1983ge,*Haber:1984rc} is widely considered to be a well-motivated possible extension of the
Standard Model that might become apparent at the TeV scale. It is therefore natural to consider supersymmetric models of
inflation. These were originally proposed because of the problems of 
the new inflationary theory~\cite{Linde:1981mu,Albrecht:1982wi} based on the one-loop (Coleman-Weinberg) potential for
breaking SU(5). Several of these problems are related to the magnitude of the effective
potential parameters: in any model of inflation based on an elementary scalar field, some parameter in
the effective potential must be small in natural units, {\textit{e.g.}}, the quartic coupling $\lambda$ in a chaotic
model with a quartic potential, or the mass parameter $\mu$ in a model of chaotic quadratic inflation.
These parameters are renormalized multiplicatively in a supersymmetric theory, so that the
quantum corrections to small values would be under control. Hence, it was
suggested that inflation cries out for supersymmetry~\cite{inflation:Cries}, though non-supersymmetric
resolutions of the problems of Coleman-Weinberg inflation are also possible: see, {\textit{e.g.}}, Ref.~\cite{Okada:2013vxa}.

\begin{pdgstrip}
\pdgwidefigure{DKLreheating_2025.pdf}
{Bayes factors calculated in~Refs.~\protect\cite{Martin:2016oyk,Martin:2024qnn} for 287 single-field
inflationary models
using \textit{Planck} 2018 data~\protect\cite{Planck:2018jri} in combination with other CMB and BAO measurements.
Those highlighted in yellow are featured in this review, according to the numbers listed in the text (the models labeled "2" and "3" lie outside the plot frame because of their low Bayesian evidence). On the vertical axis, the information gain on the reheating parameter, {\textit{i.e.}} the $k$-independent terms in the right-hand side of Eq.~\eqref{inflation:eq:nefolds}, is displayed in units of bits for the Kullback-Leibler divergence between its prior and posterior distributions~\protect\cite{Martin:2016oyk}.}
{inflation:fig:Bayes}{!htb}{width=0.9\linewidth}
\end{pdgstrip}

\index{Inflaton!Higgs scalar field as candidate}%
In the Standard Model there is only one scalar field that could be a candidate for the inflaton,
namely the Higgs field discussed above, but
even the minimal supersymmetric extension of the Standard Model (MSSM) contains many scalar fields.
However, none of these is a promising candidate for the inflaton. The minimal extension of the MSSM
that may contain a suitable candidate is the supersymmetric version of the minimal seesaw model
of neutrino masses, which contains the three supersymmetric partners of the heavy singlet (right-handed)
neutrinos. One of these singlet sneutrinos ${\tilde \nu}$ could be the inflaton~\cite{Murayama:1992ua}: it would have a quadratic
potential, the mass coefficient required would be $\sim 10^{13}$~GeV, very much in the expected ball-park
for singlet (right-handed) neutrino masses, and sneutrino inflaton decays could also give rise to the
cosmological baryon asymmetry via leptogenesis. However, as seen in \Fig{inflation:fig:Planck2018}
and already discussed, a purely quadratic inflationary potential is no longer favored by the data. This
difficulty could in principle be resolved in models with multiple sneutrinos~\cite{Ellis:2013iea}, or by postulating a trilinear
sneutrino coupling and hence a superpotential of Wess-Zumino type~\cite{Croon:2013ana}, which can yield successful
inflation with predictions intermediate between those of natural inflation and hilltop inflation in
\Fig{inflation:fig:Planck2018}.
Finally, we note that it is also possible to obtain inflation via supersymmetry breaking, as in the model of Ref.~\cite{Dvali:1994ms},
whose predictions are illustrated in \Fig{inflation:fig:Planck2018}.

\subsection{Supergravity Models}
\label{inflation:models:supergravity}

\index{Inflation models!supergravity}%

Any model of early-Universe cosmology, and specifically inflation, must necessarily incorporate gravity.
In the context of supersymmetry this requires an embedding in some supergravity theory~\cite{Nanopoulos:1982bv,Goncharov:1983mw}.
An ${\cal N} = 1$ supergravity theory
is specified by three functions: a Hermitian function of the matter scalar fields $\phi^i$, called the K\"ahler potential $K$,
that describes its geometry, a holomorphic function of the superfields, called the superpotential $W$,
which describes their interactions, and another holomorphic function $f_{\alpha \beta}$, which describes their couplings
to gauge fields $V_\alpha$~\cite{Cremmer:1982en}.

The simplest possibility is that the K\"ahler metric is flat:
\begin{equation}\label{inflation:eq:eq45}
K = \phi^i \phi_i^* \, ,
\end{equation}
where the sum is over all scalar fields in the theory, and
the simplest inflationary model in minimal supergravity has the superpotential~\cite{Holman:1984yj}
\begin{equation}\label{inflation:eq:hrrW}
W = m^2 (1 - \phi)^2 \, ,
\end{equation}
where $\phi$ is the inflaton. However, this model predicts a tilted scalar perturbation spectrum, $n_{\rm s} = 0.933$,
which is now in serious disagreement with the data from \textit{Planck} and other experiments shown in
\Fig{inflation:fig:Planck2018}.

Moreover, there is a general problem that arises in any supergravity theory coupled to matter,
namely that, since its effective scalar potential contains a factor of $e^K$, scalars typically receive squared masses
$\propto H^2 \sim V$, where $H$ is the Hubble parameter~\cite{Copeland:1994vg,*Stewart:1994ts}, 
an issue called the `$\eta$ problem'. The theory given by Eq.~(\ref{inflation:eq:hrrW})
avoids this $\eta$ problem, but a generic supergravity inflationary model encounters this problem of a
large inflaton mass. Moreover, there are additional challenges for supergravity inflation associated with the spontaneous
breaking of local supersymmetry~\cite{Coughlan:1983ci,*Goncharov:1984qm,*Banks:1993en,*deCarlos:1993wie,*Kawasaki:1995cy,Ellis:1986zt,Moroi:1994rs}.

Various approaches to the $\eta$ problem in supergravity have been proposed, including the possibility
of a shift symmetry~\cite{Kawasaki:2000yn,*Nakayama:2013txa}, and one possibility that has attracted renewed attention recently is no-scale
supergravity~\cite{Cremmer:1983bf,Goncharov:1985ka,*Kounnas:1984vh,*inflation:oldnoscale-sr2}. 
This is a form of supergravity with a K\"ahler potential that can be written in the form~\cite{Ellis:1984bm}
\begin{equation}\label{inflation:eq:noscaleK}
K \; = \; - 3 \ln \left( T + T^* - \frac{\sum_i |\phi^i|^2}{3} \right) \, ,
\end{equation}
which has the special property that it naturally has a flat potential, at the classical level and before
specifying a non-trivial superpotential. As such, no-scale supergravity is well-suited for constructing models of
inflation. Adding to its attraction is the feature that compactifications of string theory to supersymmetric
four-dimensional models yield effective supergravity theories of the no-scale type~\cite{Witten:1985xb}. There are many examples
of superpotentials that yield effective inflationary potentials for either the $T$ field (which is akin to a
modulus field in some string compactification) or a $\phi$ field (generically representing matter)
that are of the same form as the effective potential of
the $\R^2$ model Eq.~(\ref{inflation:eq:Staropot}) when the magnitude of the inflaton field is far above the Planck scale, as required to obtain
sufficiently many e-folds of inflation, $N_*$~\cite{Ellis:2013nxa,Ellis:2015xna}. This framework also offers the possibility of
using a suitable superpotential to construct models with effective potentials that are similar, but not identical, to
the $\R^2$ model, as shown by the dashed coloured lines in \Fig{inflation:fig:Staropot}.

\subsection{Other Exponential Potential Models}
\label{inflation:models:other}

This framework also offers the possibility~\cite{Ellis:2013nxa} of constructing models in which the asymptotic constant
value of the potential at large inflaton field values is approached via a different exponentially-suppressed term:
\begin{equation}\label{inflation:eq:StaroAlambdaB}
V (\phi) \; = \; A \left[ 
1 - \delta e^{-B\phi} + {\cal O}(e^{-2B\phi})
 \right] \, ,
\end{equation}
where the magnitude of the scalar density perturbations fixes $A$, but
$\delta$ and $B$ are regarded as free parameters. In the case of $\R^2$ inflation, in \Eq{inflation:eq:Staropot},
$\delta = 2$ and $B = \sqrt{2/3}$. In a general model such as \Eq{inflation:eq:StaroAlambdaB},
one finds at leading order in the small quantity $e^{-B\phi}$ that
\begin{align}
n_{\rm s} & =  1 - 2 B^2 \delta e^{-B\phi} \, , \nonumber\\
r & =  8 B^2 \delta^2 e^{-2B\phi} \, , \nonumber\\
N_* & =  \frac{1}{B^2 \delta} e^{+ B\phi} \, . \label{inflation:eq:eq46}
\end{align}
yielding the relations
\begin{equation}\label{inflation:eq:relations}
n_{\rm s} \; = \; 1 - \frac{2}{N_*} \; , \quad r \; =  \; \frac{8}{B^2 N_*^2} \, .
\end{equation}
%
These relations correspond to the class of models labeled as `$\alpha$ attractors'~\cite{Kallosh:2013yoa} in \Fig{inflation:fig:Planck2018}.
%
The spectral index in such models is determined by $N_*$, the number of e-folds Eq.~\eqref{imflation:eq:eq9} when modes exited the Hubble scale before the end of inflation.
For example, {\textit{Planck}} constraints Eq.~\eqref{inflation:eq:eq38} are consistent with $51<N_*<65$, placing bounds on the reheating temperature through Eq.~\eqref{inflation:eq:nefolds}. 
If the shift of $n_s$ towards higher values suggested by the combination with recent ACT data~\cite{ACT:2025fju} were to be confirmed, requiring values of $N_*>60$ at 2$\sigma$, the simplest realisations of $\alpha$-attractor models with conventional reheating would be ruled out~\cite{Drees:2025ngb}.

Meanwhile CMB bounds on the tensor-to-scalar ratio translate into an upper bound on the parameter $\alpha=2/3B^2$,  which determines the curvature of the K\"ahler manifold in supergravity models~\cite{Ellis:2021kad}.
There are generalizations of the simplest no-scale model \eqref{inflation:eq:noscaleK} with prefactors before the logarithmic factor
that are 1 or 2, leading to larger values of $B = \sqrt{2}$ or $1$, respectively, and hence smaller values
of $r$ than in the $\R^2$ model.
CMB constraints on the scalar spectral index can also place a lower bound on the magnitude of $\alpha$~\cite{Iacconi:2023mnw}, due to the implicit dependence of $N_*$ in Eq.~\eqref{inflation:eq:relations} upon $\alpha$~\cite{Ellis:2015pla}.


\section{Model Comparison}

Given a particular inflationary model, one can obtain constraints on the model parameters, 
informed by the likelihood, corresponding to the probability of the data given a particular 
choice of parameters (see \Sec{stat}, ``Statistics'' review).
In the light of the detailed constraints on the statistical distribution of primordial 
perturbations now inferred from high-precision observations of the CMB, it is also possible to make quantitative comparison of the statistical 
evidence for or against different inflationary models.
This can be done either by comparing the logarithm of the maximum likelihood that 
can be obtained for the data using each model, {\textit{i.e.}},
the minimum $\chi^2$
(with some correction for the number of free parameters in each model),
or by a Bayesian model comparison~\cite{Liddle:2007fy} (see also \Sec{stat:sec:bayesmod} in ``Statistics'' review).

In such a Bayesian model comparison one computes~\cite{Martin:2013tda,Martin:2013nzq,Martin:2015dha} 
the evidence, ${\cal E}({\cal D}|{\cal M}_A)$ for a model, ${\cal M}_A$, given the data ${\cal D}$. 
This corresponds to the likelihood, ${\cal L}(\theta_{Aj})=p({\cal D}|\theta_{Aj},{\cal M}_A)$, 
integrated over the assumed prior distribution, $\pi(\theta_{Aj}|{\cal M}_A)$, for all the model 
parameters $\theta_{Aj}$:
\begin{equation}\label{inflation:eq:evidence}
{\cal E}({\cal D}|{\cal M}_A) = \int {\cal L}(\theta_{Aj}) \pi(\theta_{Aj}|{\cal M}_A) d\theta_{Aj}\,.
\end{equation}
The posterior probability of the model given the data follows from 
\index{Bayes' theorem}%
Bayes' theorem
\begin{equation}\label{inflation:eq:eq47}
p({\cal M}_A|{\cal D}) = {\frac{{\cal E}({\cal D}|{\cal M}_A)\pi({\cal M}_A)}{ p({\cal D})}} \,,
\end{equation}
where the prior probability of the model is given by $\pi({\cal M}_A)$.
Assuming that all models are equally likely a priori, $\pi({\cal M}_A)=\pi({\cal M}_B)$, 
the relative probability of model $A$ relative to a reference model, in the light of the 
data, is thus given by the Bayes factor
\begin{equation}\label{inflation:eq:eq48}
B_{A,{\rm ref}} = {\frac{{\cal E}({\cal D}|{\cal M}_A)}{{\cal E}({\cal D}|{\cal M}_{\rm ref})}} \,.
\end{equation}
Computation of the multi-dimensional integral (\Eq{inflation:eq:evidence}) is a challenging 
numerical task. Even using an efficient sampling algorithm requires hundreds of thousands 
of likelihood computations for each model, though slow-roll approximations can be used 
to rapidly calculate the primordial power spectrum using the APSIC numerical 
library~\cite{2018ascl.soft06031M} for a large number 
of single-field, slow-roll inflation models.

The change in $\chi^2$ for selected slow-roll models relative to the
Starobinsky $\R^2$ inflationary model, used as a reference, is given in
\Tab{inflation:tab:inflation-evidence}
(values taken from~\cite{Martin:2024qnn}). 
\Tab{inflation:tab:inflation-evidence} also shows the Bayesian evidence 
as displayed in \Fig{inflation:fig:Bayes} for all single-field models (see Refs~\cite{Martin:2013tda, Martin:2013nzq, Martin:2024qnn} for more complete descriptions of the  models and their priors). The Jeffrey's scale distinguishes favored, weakly disfavored, moderately disfavored and strongly disfavored models, based on their Bayesian evidence. Disfavored models are those that either produce too large tensor modes, or scalar fluctuations with a spectral tilt that differs from the value preferred by the data, or that require an extreme fine tuning of their parameters.
We note that although $\alpha$-attractor models can provide a good fit to the data,
they are slightly disfavored relative to the Starobinsky model due to their larger prior volume.
There is now strong evidence ($|\ln B_{A,{\rm ref}}|>5$) against large-field models 
such as chaotic inflation with a quadratic or a quartic potential.
Indeed, over 40\% of the slow-roll inflation models considered in 
Ref.~\cite{Martin:2013tda,Martin:2013nzq,Martin:2015dha,Martin:2024qnn} are strongly 
disfavored by the \textit{Planck} data.

\begin{pdgtable}{c c c c}
{Observational evidence for and against selected inflation models: 
$\Delta\chi^2$ and the Bayes factors are calculated relative to the Starobinsky $\R^2$ 
inflationary model, which is treated as a reference. Results taken from Ref.~\protect\cite{Martin:2024qnn}.}
{inflation:tab:inflation-evidence}{}
\pdgtableheader{ Label in \Fig{inflation:fig:Bayes} & Model & $\Delta\chi^2$ & $\ln B_{A,{\rm ref}}$}
1 & $\R^2$ inflation & 0 & 0 \\
2 & Power-law potential  $\phi^2$ & $+20.76$ & $-7.36$ \\
3 & Power-law potential $\phi^4$ & $+20.81$ & $-7.34$ \\
4 &Power-law potential $\phi^{2/3}$ & $+10.82$ & $-5.49$ \\
5 &Power-law potential $\phi^{p}$ & $-0.579$ & $-4.01$ \\
6 & Hilltop potential $1-\phi^2$ & $-1.30$ & $-2.10$\\ 
7 & Hilltop potential $1-\phi^4$ & $-1.27$ & $-1.16$\\ 
8 & Hilltop potential $1-\phi^p$ & $-1.30$ & $-0.67$\\ 
9 & Brane inflation & $-1.25$ & $-0.59$ \\ 
10 & Natural inflation & $+0.55$ & $-4.74$ \\
11 & $\alpha$  attractor & $-1.26$ &$-0.28$\\
\end{pdgtable}

Models discussed in this review are highlighted in yellow in \Fig{inflation:fig:Bayes}, and numbered as follows: (1)
$\R^2$ inflation (\Sec{inflation:models:R2}) and models with similar predictions, 
such as Higgs inflation (\Sec{inflation:models:higgs})
and no-scale supergravity inflation (\Sec{inflation:models:supergravity});
chaotic inflation models (2) with a $\phi^2$ potential; (3) with a $\phi^4$ potential; (4) with a $\phi^{2/3}$ potential, and
(5) with a $\phi^p$ potential marginalising over $p \in [0.2, 6]$ (\Sec{inflation:models:chaotic}); hilltop inflation models (6) with $p = 2$;
(7) with $p = 4$ and (8) marginalising over $p$
(\Sec{inflation:models:hilltop}); (9) brane inflation
\Sec{inflation:models:brane}); (10) natural inflation
(\Sec{inflation:models:natural}); and (11) exponential potential models such as 
$\alpha$-attractors (\Sec{inflation:models:other}). 

In \Fig{inflation:fig:Bayes}, the information gain on reheating is also displayed. It corresponds to the Kullback-Leibler divergence (in units of bits) between the prior and posterior distributions of the so-called reheating parameter, formed by the right-hand side of Eq.~(\ref{inflation:eq:nefolds}). With recent CMB measurements the best inflationary models have crossed the one-bit threshold~\cite{Martin:2014nya,Martin:2016oyk}, which opens up the possibility to constrain the physical processes at play during reheating. This is what is discussed in the next Section.


\section{Constraints on Reheating}
\label{sec:reheating}

One connection between inflation and particle physics is provided by inflaton decay, whose products are expected
to have thermalized subsequently. As seen in \Eq{inflation:eq:nefolds}, the number of e-folds required during inflation depends on
details of this reheating process, including the matter density upon reheating, denoted by $\rho_{\rm th}$, which depends in turn on
the inflaton decay rate $\Gamma_\phi$. We see in \Fig{inflation:fig:Planck2018} that, within any specific inflationary model,
both $n_{\rm s}$ and $r$ are sensitive to the value of $N_*$. In particular, the uncertainty in the
experimental measurement of $n_{\rm s}$ is comparable to the variation in many model predictions for $N_* \in [50, 60]$.
This implies that the data start to constrain scenarios for inflaton decay in many models. For example, it is clear from
\Fig{inflation:fig:Planck2018} that $N_* = 60$ would be preferred over $N_* = 50$ in a chaotic inflationary model with a
quadratic potential.

As a specific example, let us consider $\alpha$-attractor models
that predict small values of $r$. 
As seen in \Fig{inflation:fig:Planck2018}, within these models
the combination of \textit{Planck}, BICEP2/Keck Array and BAO data would require a limited range of $n_{\rm s}$,
corresponding to a limited range of $N_*$, as seen by comparing the left and right vertical axes in \Fig{inflation:fig:N*plot}:
\begin{equation}\label{inflation:eq:N*limits}
N_* \; \gtrsim \; 52 \quad {\textrm{(68\%~CL)}}, \quad N_* \; \gtrsim \; 44 \quad {\textrm{(95\%~CL)}} \, .
\end{equation}
Within any specific model for inflaton decay, these bounds can be translated into constraints
on the effective decay coupling. For example, if one postulates a two-body inflaton decay
(\Eq{imflation:eq:eq14}),
the bounds (\Eq{inflation:eq:N*limits}) can be translated into bounds on 
$y^2\equiv 8\pi\Gamma/m$.
This is illustrated in \Fig{inflation:fig:N*plot}~\cite{Ellis:2021kad}, where any value of $N_{0.05}$ (the number of e-folds for a scale $0.05$~Mpc$^{-1}$, on the left vertical axis), 
corresponding to the scalar tilt $n_{\rm s}$ shown on the right vertical axis, when projected onto the
diagonal line representing one of the three illustrated choices of $\alpha$-attractor models, corresponds to a specific value of
$y$ (lower horizontal axis) and $T_{\rm reh}$ (upper horizontal axis). For example, for $\alpha = 1$ one has
\begin{equation}\label{inflation:eq:ylimits}
y \; \gtrsim \; 10^{-10} \quad {\textrm{(68\%~CL)}}, \quad y \; \gtrsim \; 10^{-17} \quad {\textrm{(95\%~CL)}} \, .
\end{equation}
These bounds are not very constraining, but they can be expected to improve significantly in
the coming years and thereby provide significant information on the connections between
inflation and particle physics.

\section{Beyond Single-Field Slow-Roll Inflation}

There are numerous possible scenarios beyond the simplest single-field models of 
slow-roll inflation. These include 
models with features in the potential leading to deviations from slow roll, as well as
theories in which non-canonical fields are 
considered, such as k-inflation~\cite{ArmendarizPicon:1999rj} or DBI 
inflation~\cite{Alishahiha:2004eh}, and multiple-field models, such as the 
curvaton scenario\cite{Enqvist:2001zp,Lyth:2001nq,Moroi:2001ct}. As well as 
altering the single-field slow-roll predictions for the primordial curvature power spectrum 
(\Eq{inflation:eq:defPzeta}) and the tensor-scalar ratio 
(\Eq{inflation:eq:singlefieldr}), they may introduce new quantities that vanish 
in single-field slow-roll models, such as isocurvature matter perturbations, 
corresponding to entropy fluctuations in the photon-to-matter ratio, at first order:
\begin{equation}\label{inflation:eq:defSm}
S_{\rm m} = {\frac{\delta n_{\rm m}}{ n_{\rm m}}} - {\frac{\delta n_\gamma}{n_\gamma}} = {\frac{\delta \rho_{\rm m}}{\rho_{\rm m}}} - {\frac{3}{4}} {\frac{\delta\rho_\gamma}{\rho_\gamma}}  \, .
\end{equation}
\index{CMB!non-Gaussianity}%
\index{Non-Gaussianity CMB}%
Another possibility is non-Gaussianity in the distribution of the primordial curvature perturbation
(see Chap. \ref{microwave}, ``Cosmic Microwave Background'' review), encoded 
in higher-order correlators such as the primordial bispectrum~\cite{Bartolo:2004if}
\begin{equation}\label{inflation:eq:3point}
    \langle \zeta({\bf k }) \zeta({\bf k^\prime}) \zeta({\bf k^{\prime \prime}}) \rangle \; \equiv \; (2 \pi)^3 \delta ({\bf k} + {\bf k^\prime} + {\bf k^{\prime \prime}})
B_\zeta(k,k^\prime,k^{\prime \prime}) \, ,
\end{equation}
which is often expressed in terms of a dimensionless non-linearity
parameter\\ $$f_{\rm NL}\propto B_\zeta(k,k^\prime,k^{\prime\prime}) / P_\zeta(k)
P_\zeta(k^\prime).$$
\noindent
The three-point function (\Eq{inflation:eq:3point}) can be thought of as defined on a triangle whose
sides are ${\bf k}, {\bf k^\prime}, {\bf k^{\prime \prime}}$, of which only two 
are independent, since they sum to zero. Further assuming statistical isotropy 
ensures that the bispectrum depends only on the magnitudes of the three vectors, 
$k$, $k^\prime$ and $k^{\prime\prime}$.
The search for $f_{\rm NL}$ and other non-Gaussian effects was a prime objective 
of the \textit{Planck} data analysis~\cite{Ade:2015ava,Akrami:2019izv}.

\subsection{Enhanced perturbations on small scales}
\label{inflation:sec:enhancements}

Although the observed primordial power spectrum on CMB scales is constrained to be almost scale invariant, consistent with the simplest slow-roll inflation models, the form of the power spectrum is largely unconstrained on much smaller scales, corresponding to scales that would have left the horizon closer to the end of inflation.
%
Although free-streaming and other damping of perturbations in the primordial plasma are expected to have washed out small-scale fluctuations in the matter sector~\cite{Chluba:2015bqa}, gravitational relics such as primordial black holes~\cite{Zeldovich:1967lct,Hawking:1974rv} or induced gravitational waves~\cite{Matarrese:1993zf,Matarrese:1997ay,Ananda:2006af,Baumann:2007zm} produced by sufficiently large primordial density perturbations could have survived and might be detectable in the present day. Rapid growth of the power spectrum on small scales requires violation of slow roll in single-field models~\cite{Motohashi:2017kbs}. This could arise from features in the scalar field potential, such as an inflection point~\cite{Garcia-Bellido:2017mdw}, leading to a transient phase of ultra-slow-roll evolution~\cite{Inoue:2001zt,Kinney:2005vj} which can give a $k^4$-rise in the scalar power spectrum on small scales~\cite{Leach:2001zf,Byrnes:2018txb}. 

The LIGO-Virgo-KAGRA collaboration has placed an upper limit on such a stochastic gravitational wave background (SGWB), conventionally expressed in terms of the effective density of gravitational waves per logarithmic frequency interval, of ${\rm \Omega}_{\rm GW}\leq5.8\times 10^{-9}$ for a scale-invariant spectrum in the frequency range $20-77$~Hz~\cite{KAGRA:2021kbb}. 
On the other hand, pulsar timing arrays have found growing evidence for a low-frequency SGWB at nano-Hertz frequencies~\cite{NANOGrav:2023gor,EPTA:2023fyk,Reardon:2023gzh,Xu:2023wog}. While this could be associated with an astrophysical background due to supermassive black hole binary systems, the NANOGrav collaboration report that a cosmological background provides a better fit to their 15-year dataset~\cite{NANOGrav:2023hvm}. This could be generated by large primordial density fluctuations on the corresponding comoving scales, $k\sim10^6$~Mpc${}^{-1}$, but these are constrained by the production of primordial black holes of approximately solar mass~\cite{Carr:2020gox} and possible one-loop or higher-order corrections to the primordial power spectrum on CMB scales. 

\subsection{Effective Field Theory of Inflation}
\label{inflation:sec:EFTinflation}

\index{Inflation!effective field theory of}%

\pdgwidefigure{N_plot2.png}
{lllustration of the impact of the BICEP/Keck~\protect\cite{BICEPKeck:2021gln} and other constraints on the inflaton decay coupling, $y$, and the number of e-folds at a scale $0.05$~Mpc$^{-1}$, $N_{0.05}$, in $\alpha$-attractor models of inflation. The horizontal blue lines are 68\% and 95\% C.L. lower limits on $n_{\rm s}$.
The left and right axes show the relation between $N_{0.05}$ and $n_{\rm s}$ and the top and bottom scales show the relation between the inflaton coupling, $y$, and the reheating temperature, $T_{\rm reh}$. The diagonal dashed, solid and dotted black lines illustrate the correlations between these quantities for $\alpha$-attractor models with $\alpha = 0.1$, $1$, and $5$. We also include the lower limit on $y$ from Big-Bang nucleosynthesis (red line), the constraint that $T_{\rm reh}$ be no smaller than the electroweak scale (gray line), and a constraint from gravitino production (purple line) for $\alpha = 1$, which strengthens for smaller $\alpha$. Plot taken from Ref.~\protect\cite{Ellis:2021kad}.}
{inflation:fig:N*plot}{!htb}{width=0.9\linewidth}

Since slow-roll inflation is a phase of accelerated expansion with an almost constant 
Hubble parameter, one may think of inflation in terms of an effective theory where the 
de Sitter spacetime symmetry is spontaneously broken down to RW symmetry by the 
time-evolution of the Hubble rate, $\dot{H}\neq0$.
%
There is then a Goldstone boson, $\pi$, associated with the spontaneous breaking 
of time-translation invariance, which can be used to study model-independent 
properties of inflation. The Goldstone boson describes a spacetime-dependent 
shift of the time coordinate,
corresponding to an adiabatic perturbation of the matter fields:
%
\begin{equation}\label{inflation:eq:eq49}
\delta\phi_i (t,\vec{x}) = \phi_i(t+\pi(t,\vec{x})) - \phi_i(t) \,.
\end{equation}
Thus adiabatic field fluctuations can be absorbed into the spatial metric perturbation, 
${\cal R}$ in \Eq{inflation:eq:Rcomoving} at first order, in the comoving gauge:
\begin{equation}\label{inflation:eq:eq50}
{\cal R} = - H\pi \, ,
\end{equation}
where we define $\pi$ on spatially-flat hypersurfaces.
In terms of inflaton field fluctuations, we can identify $\pi\equiv\delta\phi/\dot\phi$, 
but in principle this analysis is not restricted to inflation driven by scalar fields.

The low-energy effective action for $\pi$ can be obtained by writing down the most general 
Lorentz-invariant action
and expanding in terms of $\pi$. The second-order effective action for the free-field wave 
modes, $\pi_k$, to leading order in slow roll, is then
\begin{equation}\label{inflation:eq:eq51}
S^{(2)}_\pi  =  - \int d^4x \sqrt{-g} \frac{M_{\rm P}^2 \dot{H}}{c_{\rm s}^2} \left[ \dot\pi_k^2 - \frac{c_{\rm s}^2}{R^2}(\nabla\pi)^2 
 \right] \, ,
\end{equation}
where $\epsilon_H$ is the Hubble slow-roll parameter (\Eq{inflation:eq:epsilonH}).
We identify $c_{\rm s}^2$ with an effective sound speed, generalising canonical slow-roll 
inflation, which is recovered in the limit $c_{\rm s}^2\to1$.

The scalar power spectrum on super-Hubble scales (\Eq{inflation:eq:defPzeta}) is 
enhanced for a reduced sound speed, leading to a reduced tensor-scalar ratio 
(\Eq{inflation:eq:singlefieldr})
\begin{equation}\label{inflation:eq:modPzeta}
{\cal P}_\zeta (k) \simeq  \frac{4\pi}{M_{\rm P}^2} \frac{1}{c_{\rm s}^2 \epsilon} \left( \frac{H}{2\pi} \right)_*^2 \,,
\qquad
r \simeq 16 ( c_{\rm s}^2 \epsilon )_* \,.
\end{equation}
At third perturbative order and to lowest order in derivatives, one obtains\cite{Senatore:2009gt}
\begin{equation}\label{inflation:eq:eq52}
S^{(3)}_\pi  =  \int d^4x \sqrt{-g} \frac{M_{\rm P}^2(1-c_{\rm s}^2)\dot{H}}{c_{\rm s}^2} 
\left[ \frac{\dot\pi(\nabla\pi)^2}{R^2} - \left( 1+ {\frac{2}{3}} \frac{\tilde{c}_3}{c_{\rm s}^2}\right) \dot\pi^3 \right] \,.
\end{equation}
Note that this expression vanishes for canonical fields with $c_{\rm s}^2=1$. 
For $c_{\rm s}^2\neq1$ the cubic action is determined by the sound speed and an 
additional parameter $\tilde{c}_3$.
Both terms in the cubic action give rise to primordial bispectra that are 
well approximated by equilateral bispectra. However, the shapes are not 
identical, so one can find a linear combination for which the equilateral 
bispectra of each term cancel, giving rise to a distinctive orthogonal-type 
bispectrum\cite{Senatore:2009gt}.

Analysis based on \textit{Planck} 2018 temperature and polarization data has 
placed bounds on several bispectrum shapes including equilateral and orthogonal 
shapes~\cite{Akrami:2019izv}:
\begin{equation}\label{inflation:eq:eq53}
f^{\rm equil}_{\rm NL} \; = \; -26 \pm 47 
\, , 
\qquad
f^{\rm orthog}_{\rm NL} \; = \; -38 \pm 24 \quad {\textrm{(68\%~CL)}} \, .
\end{equation}
For the simplest case of a constant sound speed, and marginalising over 
$\tilde{c}_3$, this provides a bound on the inflaton sound speed~\cite{Akrami:2019izv}
\begin{equation}\label{inflation:eq:eq54}
c_{\rm s} \geq 0.021 \quad {\textrm{(95\%~CL)}} \, .
\end{equation}
For a specific model such as DBI inflation~\cite{Alishahiha:2004eh}, corresponding 
to $\tilde{c}_3=3(1-c_{\rm s}^2)/2$, one obtains a tighter bound~\cite{Akrami:2019izv}:
\begin{equation}\label{inflation:eq:eq55}
c_{\rm s}^{\rm DBI}\geq 0.086 \quad {\textrm{(95\%~CL)}} \, .
\end{equation}
The \textit{Planck} team analysed a wide range of non-Gaussian templates from 
different inflation models, including tests for deviations from an initial 
Bunch-Davies vacuum state, direction-dependent non-Gaussianity, and feature 
models with oscillatory bispectra~\cite{Akrami:2019izv}. No individual feature 
or resonance is above the $3\sigma$ significance level after accounting for 
the look-elsewhere effect. These results are consistent with the simplest canonical, 
slow-roll inflation models, but do not rule out most alternative models; rather, 
bounds on primordial non-Gaussianity place important constraints on the parameter 
space for non-canonical models.

\subsection{Multi-Field Fluctuations}

\index{Inflation!multi-field}%

There is a very large literature on two- and multi-field models of inflation, most 
of which lies beyond the
scope of this review~\cite{Gordon:2000hv,*Easther:2013rva,*Ellis:2014opa,*Turzynski:2014tza,Byrnes:2006fr}. 
However, two important general topics merit being mentioned here, namely residual isocurvature perturbations
and the possibility of non-Gaussian effects in the primordial perturbations.

One might expect that other scalar fields besides the inflaton might have non-negligible values
that evolve and fluctuate in parallel with the inflaton, without necessarily making the dominant contribution
to the energy density during the inflationary epoch. However, the energy density in such a field might persist
beyond the end of inflation before decaying, at which point it might come to 
dominate (or at least make a non-negligible contribution to) the total energy 
density. In such a case, its fluctuations could end up generating the density 
perturbations detected in the CMB. This could occur due to a late-decaying scalar 
field~\cite{Enqvist:2001zp,Lyth:2001nq,Moroi:2001ct} or a field fluctuation that 
modulates the end of inflation\cite{Lyth:2005qk} or the inflaton decay\cite{Dvali:2003em}.

\subsubsection{Isocurvature Perturbations}

Primordial perturbations arising in single-field slow-roll inflation are necessarily 
adiabatic, {\textit{i.e.}}, they affect the overall density without changing the ratios of 
different contributions, such as the photon-matter ratio, $\delta(n_\gamma/n_{\rm m})/(n_\gamma/n_{\rm m})$. 
This is because inflaton perturbations represent a local shift of the time, as 
described in section~\Sec{inflation:sec:EFTinflation}:
\begin{equation}\label{inflation:eq:eq56}
\pi = {\frac{\delta n_\gamma}{\dot{n}_\gamma}} = {\frac{\delta n_{\rm m}}{\dot{n}_{\rm m}}} \,.
\end{equation}
However, any light scalar field ({\textit{i.e.}}, one with effective mass less than the 
Hubble scale) acquires a spectrum of nearly scale-invariant perturbations during 
inflation. Fluctuations orthogonal to the inflaton in field space are decoupled 
from the inflaton at Hubble exit, but can affect the subsequent evolution of the 
density perturbation. In particular, they can give rise to local variations in the 
equation of state (non-adabatic pressure perturbations) that can alter the 
primordial curvature perturbation $\zeta$ on super-Hubble scales. Since these 
fluctuations are statistically independent of the inflaton perturbations at 
leading order in slow roll~\cite{Byrnes:2006fr}, non-adiabatic field fluctuations 
can only increase the scalar power spectrum with respect to adiabatic perturbations 
at Hubble exit, while leaving the tensor modes unaffected at first perturbative order. 
Thus, the single-field result for the tensor-scalar ratio (\Eq{inflation:eq:singlefieldr}) 
becomes an inequality~\cite{Wands:2002bn}
\begin{equation}\label{inflation:eq:multifieldr}
r \leq 16\epsilon_* \,.
\end{equation}
Hence, an observational upper bound on the tensor-to-scalar ratio does not bound the 
slow-roll parameter $\epsilon$ in multi-field models.

If all the scalar fields present during inflation eventually decay completely into 
fully thermalized radiation, these field fluctuations are converted fully into
adiabatic perturbations in the primordial plasma\cite{Weinberg:2003sw}. On the other 
hand, non-adiabatic field fluctuations can also leave behind primordial isocurvature 
perturbations (\Eq{inflation:eq:defSm}) after inflation.
In multi-field inflation models it is thus possible for non-adiabatic field 
fluctuations to generate both curvature and isocurvature perturbations leading 
to correlated primordial perturbations\cite{Langlois:1999dw}.

The amplitudes of any primordial isocurvature perturbations (\Eq{inflation:eq:defSm})
are strongly constrained by the current CMB data, especially on large angular 
scales. Using temperature and low-$\ell$ polarization data yields the following 
bound on the amplitude of cold dark matter isocurvature perturbations at scale 
$k=0.002h^{-1}$Mpc$^{-1}$ (marginalising over the correlation angle and in the 
absence of primordial tensor perturbations)~\cite{Akrami:2018odb}:
\begin{equation}\label{inflation:eq:eq57}
{\frac{{\cal P}_{S_{\rm m}}}{{\cal P}_\zeta + {\cal P}_{S_{\rm m}} }} < 
0.025 \ {\textrm{(95\%~CL)}} \, . 
\end{equation}
For fully (anti-)correlated isocurvature perturbations, corresponding to a single 
isocurvature field providing a source for both the curvature and residual isocurvature 
perturbations, the bounds become significantly tighter~\cite{Akrami:2018odb}:
\begin{align}
{\frac{{\cal P}_{S_{\rm m}}}{{\cal P}_\zeta + {\cal P}_{S_{\rm m}} }} < 0.0002 \ {\textrm{(95\%~CL), correlated}} \,,\label{inflation:eq:eq58}  \\
{\frac{{\cal P}_{S_{\rm m}}}{{\cal P}_\zeta + {\cal P}_{S_{\rm m}} }} < 0.003 \ {\textrm{(95\%~CL), anti-correlated}}  \label{inflation:eq:eq59}
\end{align}

\subsubsection{Local-Type Non-Gaussianity}

Since non-adiabatic field fluctuations in multi-field inflation may lead the to 
evolution of the primordial curvature perturbation at all orders, it becomes 
possible to generate significant non-Gaussianity in the primordial curvature 
perturbation. Non-linear evolution on super-Hubble scales leads to local-type 
non-Gaussianity, where the local integrated expansion is a non-linear function 
of the local field values during inflation, $N(\phi_i)$. While the field 
fluctuations at Hubble exit, $\delta\phi_{i*}$, are Gaussian in the slow-roll 
limit, the curvature perturbation, $\zeta=\delta N$, becomes a non-Gaussian 
distribution\cite{Lyth:2005fi}:
\begin{equation}\label{inflation:eq:eq60}
\zeta = \sum_i {\frac{\partial N}{\partial\phi_i}\delta\phi_i + {\frac{1}{2}} \sum_{i,j}} {\frac{\partial^2 N}{\partial\phi_i\partial\phi_j}} \delta\phi_i  \delta\phi_j +\ldots
\end{equation}
with non-vanishing bispectrum in the squeezed limit ($k_1\approx k_2\gg k_3$):
\begin{equation}\label{inflation:eq:eq61}
B_\zeta (k_1,k_2,k_3) \approx {\frac{12}{5}}f^{\rm local}_{\rm NL} {\frac{{\cal P}_\zeta(k_1)}{4\pi k_1^3}} {\frac{{\cal P}_\zeta(k_3)}{4\pi k_3^3}}\,,
\end{equation}
where
\begin{equation}\label{inflation:eq:localfNL}
{\frac{6}{5}}f^{\rm local}_{\rm NL} = {\frac{\sum_{i,j} {\frac{\partial^2 N}{\partial\phi_i\partial\phi_j}}}{\left( \sum_i {\frac{\partial N}{\partial\phi_i}} \right)^2}} \, .
\end{equation}
Both equilateral and orthogonal bispectra, discussed above in the context of 
generalised single-field inflation, vanish in the squeezed limit, enabling the 
three types of non-Gaussianity to be distinguished by observations, in principle.

Non-Gaussianity during multi-field inflation is highly model dependent, although 
$f^{\rm local}_{\rm NL}$ can often be smaller than unity in multi-field slow-roll 
inflation~\cite{Vernizzi:2006ve}. Scenarios where a second light field plays a 
role during or after inflation can make distinctive predictions for 
$f^{\rm local}_{\rm NL}$, such as $f^{\rm local}_{\rm NL}=-5/4$ in some 
curvaton scenarios~\cite{Lyth:2005fi,Sasaki:2006kq} or $f^{\rm local}_{\rm NL}=5$ 
in simple modulated reheating scenarios~\cite{Dvali:2003em,Dvali:2003ar}.
By contrast the constancy of $\zeta$ on super-Hubble scales in single-field 
slow-roll inflation leads to a very small non-Gaussianity
\cite{Salopek:1990re,Gangui:1993tt},
and in the squeezed limit we have the simple result
$f^{\rm local}_{\rm NL}=5(1-n_{\rm s})/12$\cite{Maldacena:2002vr,Acquaviva:2002ud}.

A combined analysis of the \textit{Planck} 2018 temperature and polarization 
data~\cite{Akrami:2019izv} yields the following range for $f^{\rm local}_{\rm NL}$
defined in (\Eq{inflation:eq:localfNL}):
\begin{equation}\label{inflation:eq:PlanckFNL}
f^{\rm local}_{\rm NL} \; = \; -1 \pm 5 \quad {\textrm{(68\%~CL)}}  \, .
\end{equation}
This sensitivity is sufficient to rule out parameter regimes giving rise to 
relatively large non-Gaussianity, but insufficient to probe 
$f^{\rm local}_{\rm NL} ={\cal O}(\epsilon)$, as expected in single-field 
models, or the range $f^{\rm local}_{\rm NL} ={\cal O}(1)$ found in the 
simplest two-field models.

Local-type primordial non-Gaussianity can also give rise to a striking 
scale-dependent bias in the distribution of collapsed dark matter halos 
and thus the galaxy distribution~\cite{Dalal:2007cu,Matarrese:2008nc}. 
Bounds from high-redshift galaxy surveys are not currently competitive with the best CMB constraints.
An analysis based on the clustering of quasars in the final data release (DR16) of the extended Baryon acoustic Oscillation Spectroscopic Survey (eBOSS)~\cite{Mueller:2021tqa} yields
\begin{equation}\label{inflation:eq:eBOSSFNL}
f^{\rm local}_{\rm NL} \; = \; -12 \pm 21 \quad {\textrm{(68\%~CL)}}  \, ,
\end{equation}
while a recent analysis based on the clustering of luminous red galaxies in imaging surveys from the 
\index{DESI -- dark energy spectroscopic instrument}%
Dark Energy Spectroscopic Instrument (DESI)~\cite{Rezaie:2023lvi} finds $-10\; < \; f^{\rm local}_{\rm NL} \; < \; 58$ at 68\%~CL.

\section{Initial Conditions and Fine-tuning}

This review is based on the assumption that the inflationary paradigm is 
valid. However, it remains the object of many criticisms 
(see, {\textit{e.g.}}, Ref.~\cite{Ijjas:2013vea}), many of them related to the 
perceived unnaturalness of the required initial conditions.

Most work on inflation is done in the context of RW cosmology, which 
assumes a high degree of symmetry, or small inhomogeneous perturbations 
(usually first order) about an RW cosmology. The isotropic RW space-time 
is an attractor for many homogeneous but anisotropic cosmologies in the 
presence of a false vacuum energy density\cite{Wald:1983ky}, or a scalar 
field with suitable self-interaction potential energy\cite{Heusler:1991ep,Kitada:1991ih}. 
However it is much harder to establish the range of highly inhomogeneous initial 
conditions that yield a successful RW Universe, at least in limited studies 
initially (see, {\textit{e.g.}}, Refs~\cite{Goldwirth:1991rj,Vachaspati:1998dy}). 
A related open question is the general nature of the pre-inflationary state 
of the inflaton and other fields that could have provided initial conditions 
suitable for inflation~\cite{Ijjas:2013vea}. They would need to have 
satisfied non-trivial homogeneity and isotropy conditions, and one may 
ask how these could have arisen, whether they are plausible, and whether 
there may be some observable signature of the pre-inflationary state.
These and other criticisms of inflation were addressed in Ref.~\cite{Chowdhury:2019otk}, 
which presented studies of the sensitivity of inflation to the initial conditions. 
Complementing these studies, there have been 
numerical relativity investigations of highly inhomogeneous initial 
conditions~\cite{East:2015ggf,Clough:2016ymm,Clough:2017efm}.
The general conclusion is that inflation is rather robust with respect to 
inhomogeneities in the initial conditions in both the scalar field profile 
and the extrinsic curvature, including large tensor perturbations.

To quantify the fine-tuning of initial conditions requires a measure in the 
space of possible cosmologies~\cite{Gibbons:1986xk}; however, it has been 
argued that some of the measures historically used to frame this problem 
are formally invalid~\cite{Schiffrin:2012zf}. 
%
It is sometimes also objected that inflationary models predict the existence 
\index{Multiverse}%
of a multiverse, and potentially a loss of predictive power~\cite{Ijjas:2014nta}, 
if it undergoes the process termed eternal 
inflation~\cite{Vilenkin:1983xq,Linde:1986fd,Goncharov:1987ir}. 
However, whether this is actually a bug or a feature remains a 
topic of debate~\cite{Guth:2013sya,Linde:2014nna}. 
The existence of the multiverse is a purely philosophical problem, 
unless it has observable consequences, {\textit{e.g.}}, in the CMB. 

One might expect signatures of any pre-inflationary state to appear 
at large angular scales, {\textit{i.e.}}, low multipoles $\ell$. Indeed, various anomalies 
have been noted in the large-scale CMB anisotropies, as also discussed in 
Chap. \ref{microwave}, the ``Cosmic Microwave Background'' review, including 
a possible suppression of the quadrupole and other very large-scale anisotropies, 
an apparent feature in the range $\ell \approx 20$ to 30, and a possible 
hemispheric asymmetry. However, none of these are highly significant statistically, 
in view of the limitations due to cosmic variance~\cite{Ade:2015lrj}. They cannot 
yet be regarded as signatures of initial conditions, the multiverse or some 
pre-inflationary dynamics, such as might emerge from string theory.

A different kind of initial condition problem, called the trans-Planckian 
problem~\cite{Martin:2000xs}, is that the perturbations now seen in the CMB 
would have had wavelengths shorter than the Planck length at the onset 
of inflation. However, under quite general and conservative assumptions 
the usual inflationary predictions are quite 
robust~\cite{Brandenberger:2000wr}, although with the possibility 
of ${\cal O}((H/m_{\rm P})^n)$ corrections that might have interesting 
signatures in the CMB~\cite{Martin:2000bv}.

When inflation was first proposed~\cite{Starobinsky:1980te,Guth:1980zm} 
there was no evidence for the existence of scalar fields or the accelerated expansion of the
Universe. The situation has changed dramatically in recent years 
with the observational evidence that the cosmic expansion is currently accelerating and with the 
discovery of a scalar particle, namely the Higgs boson (see Chap. \ref{higgs}, ``Status of Higgs boson physics'' review). 
Combined with the lack of any widely accepted alternative model for the origin of cosmic structure,
these discoveries have lent support to
the idea of a primordial accelerated expansion driven by a scalar field, {\textit{i.e.}}, cosmological inflation.
In parallel, successive CMB experiments have been consistent with generic predictions of inflationary
models, although without yet providing irrefutable evidence.
It was concluded in Ref.~\cite{Chowdhury:2019otk} that the inflationary paradigm is 
not currently in trouble. However, we note that inflation via a formally 
elementary scalar inflaton should probably only be regarded as an effective 
field theory valid at energy densities hierarchically smaller than the Planck 
scale. It should eventually be embedded in a suitable ultraviolet completion, 
on which inflationary dynamics may be our clearest window.

\section{Future Probes of Inflation}

\index{Inflation!future probes of}%

Prospective future CMB experiments, both ground- and space-based are reviewed 
in the separate PDG ``Cosmic Microwave Background'' review, Chap.~\ref{microwave}.
The main emphasis in CMB experiments in the coming years will be on
ground-based experiments providing improved measurements
of $B$-mode polarization and greater sensitivity to the tensor-to-scalar 
ratio $r$, and more precise measurements at higher $\ell$
that will constrain $n_{\rm s}$ better. As is apparent from
\Fig{inflation:fig:Planck2018} and the discussion of models such as $\R^2$ inflation, 
there is a strong incentive to reach a 5-$\sigma$ sensitivity to $r \sim 3$ to $4 \times 10^{-3}$.
This could be achieved with a moderately-sized space mission with large sky
coverage~\cite{Kogut:2011xw,*inflation:spaceCMB-sr1}, along with improvements in de-lensing and foreground measurements.
The discussion in \Sec{inflation:sec:primodialPert} (see also \Fig{inflation:fig:N*plot}), also brought out the importance of reducing
the uncertainty in $n_{\rm s}$, as a way to constrain post-inflationary reheating and the connection to
particle physics. 
CMB temperature anisotropies probe primordial density perturbations down to 
comoving scales of order $50$\,Mpc, beyond which scale secondary sources of anisotropy dominate.
CMB spectral distortions could potentially constrain the amplitude and shape 
of primordial density perturbations on comoving scales from kpc to Mpc due to 
distortions caused by the Silk damping of pressure waves in the radiation 
dominated era, before the last scattering of the CMB photons but after the 
plasma can be fully thermalised\cite{Chluba:2015bqa}. 

Improved sensitivity to non-Gaussianities is also a priority. 
In addition to CMB measurements, future large-scale structure surveys will also have roles to play as
probes into models of inflation, for which there are excellent prospects.
%
%
Current surveys such as 
\index{DESI -- dark energy spectroscopic instrument}%
DESI may reach ${\rm \Delta} f_{\rm NL}\sim4$~\cite{Font-Ribera:2013rwa} 
comparable with the \textit{Planck} sensitivity.
In the future, radio surveys such as SKA will measure large-scale structure out 
to redshift $z\sim3$\cite{Maartens:2015mra}, initially through mapping the 
intensity of the neutral hydrogen 21-cm line, and eventually through radio 
galaxy surveys that will probe local-type non-Gaussianity to $f_{\rm NL}\sim1$.

\index{Galaxy clustering}%
Galaxy clustering using DESI and 
\index{Dark energy!Euclid satellite}%
\index{Satellite!Euclid, dark energy}%
\textit{Euclid} satellite data could also 
constrain the running of the scalar tilt to a precision of ${\rm \Delta}\alpha_{\rm s}\approx0.0028$, 
a factor of 2 improvement on \textit{Planck} constraints, or a precision of $0.0016$ using 
Rubin-LSST data~\cite{Font-Ribera:2013rwa}. This might allow us to confirm, or refute, the prediction from the simplest inflationary models that running should be negative~\cite{Martin:2024nlo}.

The satellite mission 
\index{Satellite!SPHEREx, inflation}%
\index{SPHEREx -- spectro-photometer for history!of epoch reionization, satellite}%
\textit{SPHEREx}~\cite{Dore:2014cca} 
will use measurements of the galaxy power spectrum to target an estimate of the running 
of the scalar spectral index with a sensitivity ${\rm \Delta}\alpha_{\rm s}\sim 10^{-3}$ and local-type 
primordial non-Gaussianity, ${\rm \Delta} f_{\rm NL}\sim1$. Including information from the galaxy 
bispectrum one might reduce the measurement error on non-Gaussianity to 
${\rm \Delta} f_{\rm NL}\sim0.2$, making it possible to distinguish between single-field 
slow-roll models and alternatives such as the curvaton scenario for the origin 
of structure, which generate $f_{\rm NL}\sim1$.
%

Finally we note that future gravitational wave experiments such as the Laser Interferometer Space Antenna (LISA) will be sensitive to a stochastic gravitational wave background, including tensor perturbations produced during inflation or induced by large primordial density perturbations on small scales after inflation~\cite{LISACosmologyWorkingGroup:2022jok}, which could provide constraints on inflationary perturbations on smaller scales.

\noindent {\bf Acknowledgements}

The work of J.E.~was supported in part by the UK STFC research grant ST/T000759/1.
The work of D.W.~was supported in part by the UK STFC research grant ST/W001225/1.

\IfFileExists{inflation.bib}{\putbib[inflation]}{}